\documentstyle[floats,prd,aps,preprint,epsf]{revtex}

\draft
\begin{document}

\newcommand{\beq}{\begin{equation}}
\newcommand{\eeq}{\end{equation}}
\newcommand{\beqn}{\begin{eqnarray}}
\newcommand{\eeqn}{\end{eqnarray}}
\newcommand{\pa}{\partial}
\newcommand{\vp}{\varphi}
\newcommand{\varep}{\varepsilon}
\def\zero{\hbox{$_{(0)}$}}
\def\bL{\hbox{$\,{\cal L}\!\!\!$--}}
\def\bI{\hbox{$\,I\!\!\!$--}}
\def\bm#1{{\hbox{\boldmath $#1$}}}
\def\riso{\hbox{$r_{\rm ISCO}$}}
\def\jisom{\hbox{$j_{\rm ISCO:max}$}}

\begin{center}
{\large\bf{New criterion for direct black hole formation in rapidly
 rotating stellar collapse}}
~\\
~\\
Yu-ichirou Sekiguchi and Masaru Shibata\\
{\em Graduate School of Arts and Sciences,
University of Tokyo, Tokyo, 153-8902, Japan}\\
\end{center}
\begin{abstract}
We study  gravitational collapse of rapidly rotating
relativistic polytropes of the adiabatic index
$\Gamma = 1.5$ and $2$, in which 
 the spin parameter $q \equiv J/M^{2} > 1$ where $J$ and $M$ are total
 angular momentum and gravitational mass, in full general relativity. 
 First, analyzing initial distributions of the mass and the
 spin parameter inside stars,  we predict the final outcome
 after the collapse. Then, we perform fully general relativistic
 simulations on assumption of axial and equatorial symmetries and 
 confirm our predictions. As a result of simulations, we find that in
 contrast with the previous belief, even for stars with $q > 1$, the
 collapse proceeds to form a seed black hole at central region, and the seed
 black hole subsequently grows as the ambient fluids accrete onto it. 
We also find that growth of angular momentum and mass
of the seed black hole can be approximately determined
from the initial profiles of the density and
the specific angular momentum.  We define an 
effective spin parameter at the central region of the stars, $q_{c}$,
and propose a new criterion for black hole formation as $q_{c} \alt 1$.
Plausible reasons for the discrepancy between our and previous
results are clarified. 
\end{abstract}
\pacs{04.25.Dm, 04.30.-w, 04.40.Dg}

%
%
\section{INTRODUCTION}\label{intro}

One of the fundamental issues in numerical general relativity is to
explore the final fate after gravitational collapse of rotating stellar cores.
If their mass is not as large as the maximum mass of a neutron star
$\sim 2 M_{\odot}$, a rotating neutron star will be the outcome of
the collapse. On the other hand, 
for stars of sufficiently large mass,
the collapse will proceed completely and
a spacetime singularity will be formed according to the singularity
theorems of Hawking and Penrose\cite{Singu0,Singu1}. 
If the cosmic censorship conjecture suggested by Penrose\cite{Pen} is
correct, any singularity should be surrounded by an event horizon. 
Then the black hole uniqueness theorems of Israel\cite{Isr},
Carter\cite{Car}, and Robinson\cite{Rob} tell that a collapsed star will 
consequently settle down to a Kerr black hole. 
On the other hand, if this conjecture is not correct, a rotating stellar
collapse might form a state with naked singularities.

It is well known that in a Kerr spacetime, 
the singularity is covered by an event horizon only if a non-dimensional
spin parameter defined as $q \equiv J/M^{2}$, 
where $J$ and $M$ are the angular momentum and the gravitational mass of 
the system, does not exceed unity\cite{Singu0,Singu1}. 
Otherwise the singularity is naked.
This implies that the value of $q$ for
any black hole cannot be larger than unity. 
For the realistic progenitor of black holes, 
however, the value of $q$ may be larger than unity
(see Sec. \ref{Candidates}). 
Thus, it is interesting to explore the final fate after gravitational collapse
of rotating stars with $q>1$.
Numerical relativity is the unique approach to resolve this problem.

There have been several studies in full general relativity 
with regard to the above
subject on the assumption of
axial symmetry\cite{Nak,NOK,SP,ACST,Shiba2000,S03,SS}. 
A series of simulations of rotating stellar collapse 
in full general relativity was first 
performed by Nakamura\cite{Nak} and his collaborators\cite{NOK} 
using the (2+1)+1 formalism developed by Maeda {\it et al}.\cite{MSNM}. 
They adopted differentially rotating massive stars that are to collapse 
(i.e., whose masses are much larger than the maximum allowed mass 
for a neutron star formation). 
An interesting finding in their simulations is that $q$ is
an important parameter for determining the prompt black hole formation. 
Their results suggest that for $q>1$, no black hole is formed and 
the stars bounce back due to centrifugal force, 
indicating that the cosmic censorship conjecture holds. 
Stark and Piran \cite{SP} performed simulations for 
the collapse of polytropes with the adiabatic index $\Gamma=2$
and with an artificially given rigid rotation, 
using the Bardeen-Piran formalism \cite{BP}. They reconfirmed 
the result found by Nakamura; $q \sim 1$ is the demarcation between
black hole formation and bounce. 
Abrahams {\it et al}.\cite{ACST} studied the collapse of axisymmetric tori
consisting of collisionless matter, and also found that black holes form
only from initial configurations with $q \alt 1$. 
Shibata \cite{Shiba2000} investigated the effects of rotation and shock 
heating on the criteria for prompt formation of black holes, for rotating 
equilibrium polytropes with $\Gamma = 2$ and $q < 1$, and 
found that effects of rotation and shock heating raise the critical
mass for black hole formation. 
Shibata \cite{S03} also studied formation of black holes from 
marginally stable supramassive rotating neutron stars, modeled by
$\Gamma = 2.5$, $2.25$, $5/3$, and $1.5$ polytropes rotating at the mass
shedding limit for which $q < 1$, and 
found that the final state of such collapse 
is a Kerr black hole with no appreciable disk. Shibata and 
Shapiro \cite{SS} studied the collapse of a 
rigidly rotating polytrope at the mass-shedding limit with $\Gamma=4/3$.
The value of $q$ for such a configuration is close to unity as
$\approx 0.96$ \cite{BS}. 
They indicate that the final state is a black hole surrounded
by an appreciable disk ($M_{\rm disk}/M \approx 0.1$).
All these results seem to suggest that 
$q \approx 1$ is the maximum value for the black hole formation,
and the final state after the gravitational collapse of rotating stars with
$q < 1$ is a rotating Kerr black hole with a small disk mass. 

However, the rotating stellar collapse in general relativity
has been not sufficiently
studied for {\it realistic} stellar cores; e.g., iron stellar 
cores and pair-unstable oxygen cores for which $\Gamma \alt 4/3$ at
the onset of the collapse. 
Recently, Shibata \cite{ShibaSA} determined 
marginally stable and rigidly rotating stars with soft equations of
state ($\Gamma \alt 4/3$) against gravitational collapse, which
are plausible initial conditions for rotating stellar core collapse.
He indicates that (i) even for a rigidly rotating star with $q > 1$, 
a central region, in which an approximate {\it local} value of $q$ ($q_c$) 
is smaller than unity, will first collapse to form a black hole,
unless $q$ is not too large ($q < 2.5$), and that (ii) 
as a result of the collapse for $q_c < 1$ and $q \gg 1$,
a massive disk ($M_{\rm disk}/M=O(0.1)$) will be formed around the
final black hole.

In this paper, we examine the validity of the conjectures 
indicated in \cite{ShibaSA} by axisymmetric simulations 
in full general relativity. 
The essence of the points in \cite{ShibaSA} is that the density profile
for the equilibrium stars with $\Gamma \alt 4/3$ is significantly
different from that for stiff equations of state, only for which
previous works have paid attention \cite{NOK,SP,Shiba2000,S03}.
This fact could modify the criterion of black hole formation
from previous studies. As a first step toward a more realistic
simulation, in this paper, 
we focus on extracting a physical essence for the criterion of
black hole formation using simple toy models. 
We perform simulations for 
rotating stellar collapse with a {\it moderately} soft equations of
state of $\Gamma=1.5$ and with a stiff equation of state
of $\Gamma=2$ and compare the results.
From the results of these simulations, 
we illustrate that the global value of $q$ 
is an adequate parameter for predicting the black hole formation only for
the stiff equations of state but inadequate for the soft equations of state. 

The paper is organized as
follows. In Sec. II, we briefly describe our formulation, gauge
conditions, and boundary conditions. In Sec. III, we first review
plausible initial conditions for black hole formation in
which the value of $q$ can be larger than unity. Then, we describe
initial conditions adopted in this paper and 
predict the final outcomes of the stellar collapse for our
initial models following \cite{ShibaSA}. Sec. IV presents
numerical results, emphasizing that the predictions made in Sec. III
are correct. Sec. V is devoted to a summary. Throughout this 
paper, we adopt the geometrical units $G = c = 1$, 
where $G$ and $c$ denote the gravitational constant and the speed of
light, respectively.
We use the Cartesian coordinates $x^{k} = (\, x, y, z\,)$  as the
spatial coordinates, with $r = \sqrt{x^{2} + y^{2} + z^{2}}$. 

\section{SUMMARY OF FORMULATION}

We perform fully general relativistic simulations for rotating 
stellar collapse in axial symmetry using 
the same formulation as 
in~\cite{S2003}, to which the reader may refer for details and basic 
equations. The fundamental variables for the hydrodynamics are: 
\beqn 
\rho &&:{\rm rest~mass~ density},\nonumber \\
\varep &&: {\rm specific~ internal~ energy}, \nonumber \\
P &&:{\rm pressure}, \nonumber \\
u^{\mu} &&: {\rm four~ velocity}, \nonumber \\
v^i &&={dx^i \over dt}={u^i \over u^t},
\eeqn
where subscripts $i, j, k, \cdots$ denote $x, y$, and $z$, and 
$\mu$ the spacetime components. 
As the variables to be evolved in the numerical simulations, 
we define a weighted density $\rho_* = \rho \alpha u^t e^{6\phi}$ 
and a weighted four-velocity
$\hat u_i = h u_i= (1+\varepsilon+P/\rho)u_i$ where $h$ denotes a 
specific enthalpy. 
{}From these variables, the total baryon rest mass and angular momentum 
of system, which are conserved quantities in axisymmetric
spacetime, can be defined as 
\beqn 
M_*&=&\int d^3 x \rho_*, \\
J  &=&\int d^3 x \rho_*\hat u_{\varphi} . 
\eeqn
General relativistic hydrodynamic equations are solved using 
a so-called high-resolution shock-capturing scheme \cite{Font,S2003} 
on the $y=0$ plane with the cylindrical coordinates $(x, z)$
(in the Cartesian coordinates with $y=0$). 

To model initial conditions we adopt the polytropic equations of state 
\beq \label{poly-EOS}
P =  K \rho^{1+\frac{1}{n}},
\eeq
where $n$ is the polytropic index and $K$ polytropic constant. 
Then physical units enter the problem only through the 
polytropic constant $K$, which can be chosen arbitrarily or else 
completely scaled out of the problem. Thus, in the following 
we display only the dimensionless quantities which are defined as
\beq \label{rel-nondim}
\bar{M}_{\ast} = M_{\ast}K^{-n/2}, \ \ \bar{M} = M K^{-n/2}, \ \ \bar{R}
= R K^{-n/2}, \ \ \bar{J} = J K^{-n}, \ \ \bar{\rho} = \rho K^{n}, \ \
\bar{\Omega} = \Omega K^{n},
\eeq
where $M$, $R$, and $\Omega$ denote an ADM mass, a radius, and an angular
velocity. Hereafter, we adopt the units of $K = 1$ so that we omit 
the bar. 

On the other hand, during the time evolution,
we use the so-called $\Gamma$-law equations of state of the form
\beq
P = (\Gamma -1)\rho \epsilon,
\eeq
where the adiabatic index $\Gamma$ is set as $1+ 1/n$. 
In the absence of shocks, no heat is generated and
the collapse proceeds in an adiabatic manner, preserving the polytropic
form of the equations of state.

We neglect effects of viscosity and magnetic fields. The timescale of dissipation and
angular momentum transport due to these effects are much longer than the
dynamical timescale, unless the magnitude of viscosity or magnetic
fields is extremely large\cite{BSS}. Thus
neglecting them is an appropriate assumption.

The fundamental variables for geometry are: 
\beqn
\alpha &&: {\rm lapse~function}, \nonumber \\
\beta^k &&: {\rm shift~vector}, \nonumber \\
\gamma_{ij} &&:{\rm metric~in~3D~spatial~hypersurface},\nonumber \\ 
\gamma &&=e^{12\phi}={\rm det}(\gamma_{ij}), \nonumber \\
\tilde \gamma_{ij}&&=e^{-4\phi}\gamma_{ij}, \nonumber \\
K_{ij} &&:{\rm extrinsic~curvature}. 
\eeqn
As in the series of our papers, we evolve $\tilde \gamma_{ij}$, $\phi$, 
$\tilde A_{ij} \equiv e^{-4\phi}(K_{ij}-\gamma_{ij} K_k^{~k})$,
and trace of the extrinsic curvature $K_k^{~k}$ 
together with three auxiliary functions
$F_i\equiv \delta^{jk}\pa_{j} \tilde \gamma_{ik}$ with an
unconstrained free evolution code
as in \cite{AMS,SBS,bina1,bina2,SN,Shiba2000,S2003}. 

The Einstein equations are solved in the Cartesian coordinates. 
To impose axisymmetric boundary conditions, the Cartoon method
is used \cite{alcu}: Assuming the reflection symmetry with 
respect to the equatorial plane, simulations are performed 
using a fixed uniform grid with the grid 
size $N \times 3 \times N$ in $(x, y, z)$ which covers 
a computational domain as 
$0 \leq x \leq L$, $0 \leq z \leq L$, and $-\Delta \leq y \leq \Delta$. 
Here, $N$ and $L$ are constants and $\Delta = L/N$. 
In the Cartoon method, the axisymmetric boundary conditions are
imposed at $y= \pm \Delta$. Details are described in \cite{S2003} to
which the readers may refer.

As the slicing condition we impose 
an ``approximate'' maximal slicing condition in which 
$K_k^{~k} \approx 0$ is required\cite{AMS}. 
As the spatial gauge, 
we adopt a dynamical gauge condition \cite{DynGauge}. In the present work,
the equation for the shift vector is written as 
\beq
\pa_t \beta^k = \tilde \gamma^{kl} (F_l +\Delta t \pa_t F_l),
\label{dyn}
\eeq
where $\Delta t$ denotes a time step in numerical computation\cite{S03}. 
Note that in this gauge condition, $\beta^{i}$ obeys a hyperbolic-type
equation for sufficiently small value of $\Delta t$ because the
right-hand side of the evolution equation for $F_i$
contains a vector Laplacian term \cite{SN}. 
The outstanding merit of this gauge condition is that we can save
computational time significantly. It has already been found that stable
simulations for rotating stellar collapse and merger of binary neutron
stars are feasible in this gauge \cite{bina2,S03}.

An outgoing wave boundary condition 
is imposed for $h_{ij}(\equiv \tilde \gamma_{ij}-\delta_{ij})$, 
$\tilde A_{ij}$, and $F_{i}$
at the outer boundaries of the computational domain. 
The condition adopted is the same as that described in \cite{SN}. 
On the other hand, for $\phi$ and $K_k^{~k}$,
outer boundary conditions are imposed as
$r\, \phi = {\rm const}$ and $K_k^{~k}=0$, respectively.

A black hole may be formed as a result of
collapse. We determine the location of it using an apparent horizon
finder developed in \cite{ShibaAH}. As the system approaches a stationary
state, the apparent horizon will approach the event horizon. In
a dynamical spacetime we compute the apparent horizon mass $M_{\rm AH}$
which is defined as\cite{YP}
\beq
M_{\rm AH} = \sqrt{\frac{A}{16\pi}} ,
\eeq
where $A$ denotes area of an apparent horizon.

During the numerical simulations,  conservation of 
mass and angular momentum and a violation of the Hamiltonian constraint, 
$H_{\rm error}$, are monitored as code checks. Here $H_{\rm error}$ is
evaluated with a weighted average by $\rho_*$ as 
\begin{equation}
H_{\rm error} \equiv \frac{\displaystyle \int \rho_{\ast}
\left| V \right| d^{3}x}
{\displaystyle \int \rho_{\ast} d^{3}x} = 
\frac{1}{M_{\ast}} \int \rho_{\ast} \left| V \right| d^{3}x,
\end{equation}
where $V$ is defined as
\begin{equation}
V \equiv \frac{\displaystyle \; \tilde \Delta \psi
- \frac{1}{8}\psi \tilde{R} + 
2\pi \psi^{5} \left(\rho h w^{2} - P \right) 
+ \frac{\psi^{5}}{8}\tilde{A}_{ij}\tilde{A}^{ij} 
- \frac{\psi^{5}}{12}(K_k^{~k})^{2}\; }
{\displaystyle \; \left|\tilde \Delta \psi\right| 
+ \frac{1}{8}\left|\psi\tilde{R}\right| 
+ 2\pi \psi^{5} \left(\rho h w^{2} - P \right) 
+ \frac{\psi^{5}}{8}\tilde{A}_{ij}\tilde{A}^{ij} 
+ \frac{\psi^{5}}{12}(K_k^{~k})^{2} \; }. \label{eq11}
\end{equation}
In Eq. (\ref{eq11}), $\psi \equiv e^{4\phi}$, $w \equiv \alpha u^t$,
and $\tilde \Delta$ and 
$\tilde{R}$ are the Laplacian and the Ricci scalar with 
respect to the conformal metric $\tilde{\gamma}_{ij}$.
Numerical results for several test
calculations, including stability and collapse of spherical
and rotating neutron stars, have been described in~\cite{S2003}. 

\section{INITIAL CONDITIONS AND PREDICTIONS}

\subsection{Candidates for black hole progenitor}\label{Candidates}

In this subsection
we review two realistic candidates for the progenitor of black holes
and recall that the value of $q$ for them may be larger than unity.

The first candidate is a degenerate iron core in excess of the
Chandrasekhar mass\cite{ST,WHW}. 
It is well known that stars whose initial mass are larger than 
$\sim 8M_{\odot}$ evolve to form a core mainly composed of iron group elements.
Because of iron being the most stable nuclei and no energy generation
by nuclear burning, the core contracts gradually. Accordingly the central
temperature, $T_{c}$, 
and central density, $\rho_{c}$, rise to be $T_{c}>10^{9}$ K and
$\rho_{c} > 10^{9}$ g$/$cm$^{3}$. Then for the stars of initial mass
larger than $\sim 10 M_{\odot}$, the photo-dissociation of iron to lighter
elements occurs first at such hot and dense environments. The entropy of
electrons is reduced, as density increases, by the loss of thermal
energy due to the photo-dissociation.  As a result, the adiabatic index
$\Gamma$ decreases below $4/3$ and the core is
destabilized\cite{ST,WHW}. If mass of the core is much larger than the
maximum neutron star mass, a black hole will be formed after the collapse.

The second candidate is 
a star of very low metallicity or a member of the so-called first stars.
Recent numerical simulations of the collapse of primordial
molecular clouds have suggested that the first generation of stars
contains many massive members of mass 
$M > 100M_{\odot}$\cite{Abel,NU,Bromm}. Such massive
stars may evolve to form large oxygen core after the helium burning
phase\cite{BAC}. If the oxygen core is sufficiently massive, it will
encounter electron-positron pair creation. As a result, thermal energy that
might have gone into raising the temperature and providing more pressure
support is diverted to it, 
bringing its adiabatic index $\Gamma$ below $4/3$ to destabilize the core. 
Following the pair creation, the oxygen-to-silicon and
silicon-to-nickel transformation will supply internal energy to the
core, which could force back the collapse to explode. However, if
mass of the core is sufficiently large, the contraction proceeds and
then nickel-to-alpha phase transition will set in,  
nullifying all decelerating effect of
the nuclear burnings. Eventually the core will completely collapse to
form a black hole \cite{BAC,FWH}.

The gravitational collapse sets in when stability of the progenitors changes
and, hence, a marginally stable star against gravitational collapse may
be regarded as an initial condition for the collapse\cite{ShibaSA}. 
Therefore, for both candidates described above, 
 initial conditions for black hole
formation are marginally stable rotating stars with adiabatic constant
$\Gamma < 4/3$. Note that for a configuration with $\Gamma < 4/3$, rotation can
stabilize it against gravitational collapse. 

Indeed, using polytropic equations of state, 
a stability criterion
against gravitational collapse can be written, including rotational effect
and post-Newtonian correction as \cite{ST,ShibaSA}
\beq \label{beta-criterion}
3\Gamma - 4 - 2(3\Gamma -5)\beta - k(\Gamma)\frac{M}{R} < 0, \ \
{\rm for}\ \ {\rm instability}, 
\eeq
where $\beta$ is the ratio of rotational kinetic energy, $T$, to 
gravitational potential energy, $W$, as
\beq
\beta \equiv \frac{T}{|W|}.
\eeq
$k$ is a parameter which is $\approx 6.75$ in the case 
$\Gamma = 4/3$ and $R$ is a radius of spherical polytrope. 
For the progenitor of black holes described above, the compactness
$M/R$ is lager than $10^{3}$ and, hence, the general relativistic
correction may be neglected. To analyze the
effect of rotation, we restrict our attention to the rigidly rotating
case for simplicity in the following. Based on a numerical analysis in
Newtonian gravity, the maximum value of $\beta$
for rigidly rotating stars, which is achieved when
the velocity at the equatorial surface is equal to the Keplerian
velocity (i.e., at the mass-shedding limit),
with polytropic equations of state is
approximately written as\cite{ShibaSA}
\beq \label{beta-max}
\beta_{\rm max} \approx 0.00902 
+ 0.124\left(\Gamma - \frac{4}{3} \right).
\eeq
Combining Eqs. (\ref{beta-criterion}) and (\ref{beta-max}), it is
found that stability criterion becomes as follows: 
\beq
\Gamma \alt 1.328, \ \
{\rm for}\ \ {\rm instability}.
\eeq
Therefore, the rotation stabilizes the star with 
$1.328 \alt \Gamma < 4/3$. This is a crucial result.
In a rapidly and 
rigidly rotating equilibrium state, the spin parameter is related to 
the compactness parameter $M/R$ as 
\beq
q \equiv \frac{J}{M^{2}} \propto \frac{M R^{2} \Omega}{M^{2}} \propto
\sqrt{\frac{R}{M}}.
\eeq
On the other hand, the value of $R/M$ for a marginally stable
star becomes larger for the smaller value of $\Gamma$ \cite{ShibaSA}, 
accordingly increasing the value of $q$. 
Recall that the value of $q$ is 
close to unity for a maximally rotating and marginally stable star
with $\Gamma = 4/3$ \cite{BS}. Taking all these into
account, it is likely that the value of $q$ exceeds unity for a
rapidly rotating and marginally stable star 
with $\Gamma$ between $1.328$ and $4/3$. 
Actually, it has been shown that this is the case for
marginally stable and rigidly 
rotating polytropes with $1.328 \alt \Gamma \alt 1.332$ \cite{ShibaSA}.
Thus, the value of $q$ for the rapidly rotating progenitor of
black hole formation {\it in nature} is likely to be larger than unity.

\subsection{Method of preparing initial conditions}\label{PrepareInitial}

\begin{table}[bt]
 \begin{center}
  \begin{tabular}{cccccccccc}
   $\rho_{c}$ & $\Omega_{0}$ & $R_{0}/R_S$ & $M$ & $M_{\ast}$ & $J$ &
   $q$ & $q_{\ast}$ & $q_c$ & $f_{P}$ \\ \hline
   $0.005$ & $0.050$ & $\infty$ (rigid) & 0.510 &
   0.530  & $0.227$ & 0.87 & 0.81 & 0.50 & 0.35 \\
   \hline
   $0.005$ & $0.065$ & $\infty$ (rigid) & 0.512 &
   0.535  & $0.300$ & 1.14 & 1.05 &0.64 & 0.25 \\
   $0.005$ & $0.065$ & $1.5$ & 0.501 & 0.534 &
   0.282 & 1.08 & 0.99 &0.64& 0.25 
   \\ \hline
   $0.005$ & $0.090$ & $\infty$ (rigid) & 0.529 &
   0.546 & 0.444 & 1.59 & 1.49 &0.84& 0.1 \\
   $0.005$ & $0.090$ & $1.0$ & 0.519 & 0.541 & 0.377 
   & 1.40 & 1.29 &0.86& 0.1 \\
   $0.005$ & $0.090$ & $2/3$ & 0.511 & 0.537 &
   0.319 & 1.22 & 1.11 &0.87& 0.1 \\ 
   \hline
   $0.005$ & $0.100$ & $\infty$ (rigid) & $0.538$
   & $0.552$ & $0.512$ & $1.77$ & 1.68 &0.91& 0.01 \\
   $0.005$ & $0.100$ & $1.0$ & $0.524$ & $0.545$ & $0.427$ 
   & $1.55$ & 1.44 &0.94& 0.01 \\
   $0.005$ & $0.100$ & $2/3$ & $ 0.514$ & $0.540$ 
   & $0.358$ & $1.36$ & 1.23 &0.96& 0.01 \\ 
   \hline
   $0.005$ & $0.115$ & $\infty$ (rigid) & $0.562$
   & $0.563$ & $0.640$  & $2.03$ & 2.02 &1.01& 0.01 \\
   $0.005$ & $0.115$ & $1.0$ & $0.540$ & $0.553$ & $0.515$ 
   & $1.77$ & 1.68 &1.05& 0.01 \\
   $0.005$ & $0.115$ & $2/3$ & $0.526$ & $0.546$ 
   & $0.319$ & $1.54$ & 1.42 &1.08& 0.01\\ 
  \end{tabular}
 \end{center}
\caption{Central density $\rho_{c}$, 
angular velocity of rotation axis $\Omega_{0}$, 
differential rotation parameter  $R_{0}$ 
($R_{S}$ is the coordinate radius of the spherical polytrope in 
the isotropic coordinates), 
gravitational mass $M$, baryon rest mass $M_{\ast}$, angular momentum $J$, 
the non-dimensional parameter (spin parameter) $q=J/M^{2}$, 
$q_{\ast}=J/M_{\ast}^{2}$, and $q_c$ of the initial data
for models with $\Gamma = 1.5$ in units of $G=c=K=1$. $f_{P}$ denotes 
a pressure reduction parameter. ``rigid'' implies the
rigid rotation model. 
}\label{Table1}
\end{table}
\begin{table}[bt]
 \begin{center}
  \begin{tabular}{cccccccccc}
   $\rho_{c}$ & $\Omega_{0}$ & $R_{0}$ & $M$ & $M_{\ast}$ & $J$ &
   $q$ & $q_{\ast}$ & $q_c$ & $f_{P}$ \\ \hline
   0.318 & 0.65 & $\infty$ (rigid) & 0.179 & 0.197 & 0.0364 &
   1.13 & 0.93 & 0.90 & 0.01\\
   0.318 & 0.67 & $\infty$ (rigid) & 0.182 & 0.199 & 0.0383 &
   1.16 & 0.97 & 0.91 & 0.01\\
   0.318 & 0.68 & $\infty$ (rigid) & 0.183 & 0.199 & 0.0393 &
   1.17 & 0.99 & 0.92 & 0.01\\
  \end{tabular}
 \end{center}
\caption{The same as \ref{Table1} but for models with 
$\Gamma = 2.0$.}\label{Table1a}
\end{table}
We are most interested in the final fate after gravitational collapse
of rotating marginally stable stars with $1.328 \alt \Gamma \alt 1.332$
and $q > 1$. Such simulation is computationally challenging because
the dynamical scale changes by a factor of $\sim 10^4$ during the
collapse and also because it is necessary to take into account
a realistic equation of state for getting scientific results. 
Fortunately, the physics that we want to understand (i.e.,
criterion for black hole formation) 
can be extracted even from a simulation with appropriate ``toy'' models.
Here, we adopt simplified initial conditions following
Stark and Piran \cite{SP}: We first give a marginally stable
spherical polytrope, and then, add an angular momentum
artificially as well as reduce the pressure to induce the collapse. 
We choose two values of $\Gamma$ as $1.5$, with which 
an insight into the collapse of stars with soft equation of state may be
gained, and $\Gamma = 2$, the value that Stark and Piran adopted. 

In more detail, we prepare initial conditions in the following
procedure. 
First, we give a spherical star, marginally stable against
gravitational collapse, using 
the polytropic equations of state with $\Gamma = 1.5$ or $2.0$. 
Second, we reduce the pressure by an arbitrarily chosen
fraction $f_{P}$ of its equilibrium pressure. 
Third, an angular momentum is artificially added 
according to 
\beq
u_{\varphi} = e^{4\phi} \varpi^{2} u^{t}\Omega ,\ \ \ {\rm and}\ \ \ 
u^{t} = \sqrt{\frac{-1}{-\alpha^{2} + \Omega^{2}e^{4\phi} \varpi^{2}}},
\eeq
where $\alpha$ and $\phi$ denote the lapse function and the conformal
factor of the spherical polytrope, and $\Omega$ is given by 
\beq
\Omega (\varpi)
= \Omega_{0}\exp \left[-\frac{\varpi^{2}}{2 R_{0}^{2}}\right] .
\eeq
Here, $\varpi = \sqrt{x^{2} + y^{2}}$ and 
$R_{0}$ is a parameter which controls the degree of differential
rotation. For $R_{0} \rightarrow \infty$, the rotation approaches the
rigid rotation (in the following we refer the case
with $R_{0} \rightarrow \infty$ as the rigid rotation case). 
Forth, the Hamiltonian and 
momentum constraints are reimposed by solving the constraint equations,
and then the time evolution is set out.

In Tables \ref{Table1} and \ref{Table1a}, we list characteristic
quantities for our initial conditions with $\Gamma =1.5$ and 
$2.0$ in the present work. $R_{S} = 6.45$ is a coordinate
radius of a marginally stable spherical polytrope with $\Gamma = 1.5$ in
the isotropic coordinates and all the quantities are scaled to be
non-dimensional using the relations (\ref{rel-nondim}).

\subsection{Predicting the final outcome}\label{Prediction}

\begin{figure}[tb]
\begin{center}
\epsfxsize=3.3in
\leavevmode
(a)\epsffile{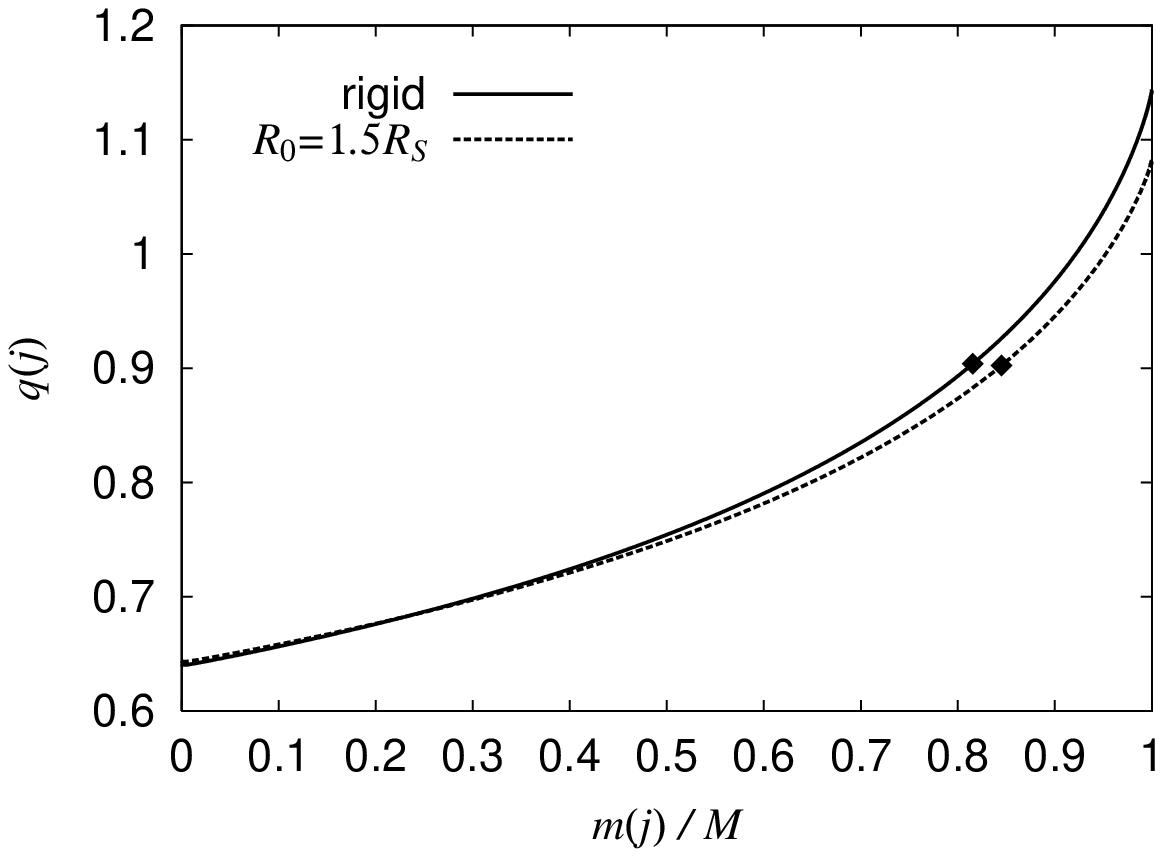} \\
\epsfxsize=3.3in
\leavevmode
(b)\epsffile{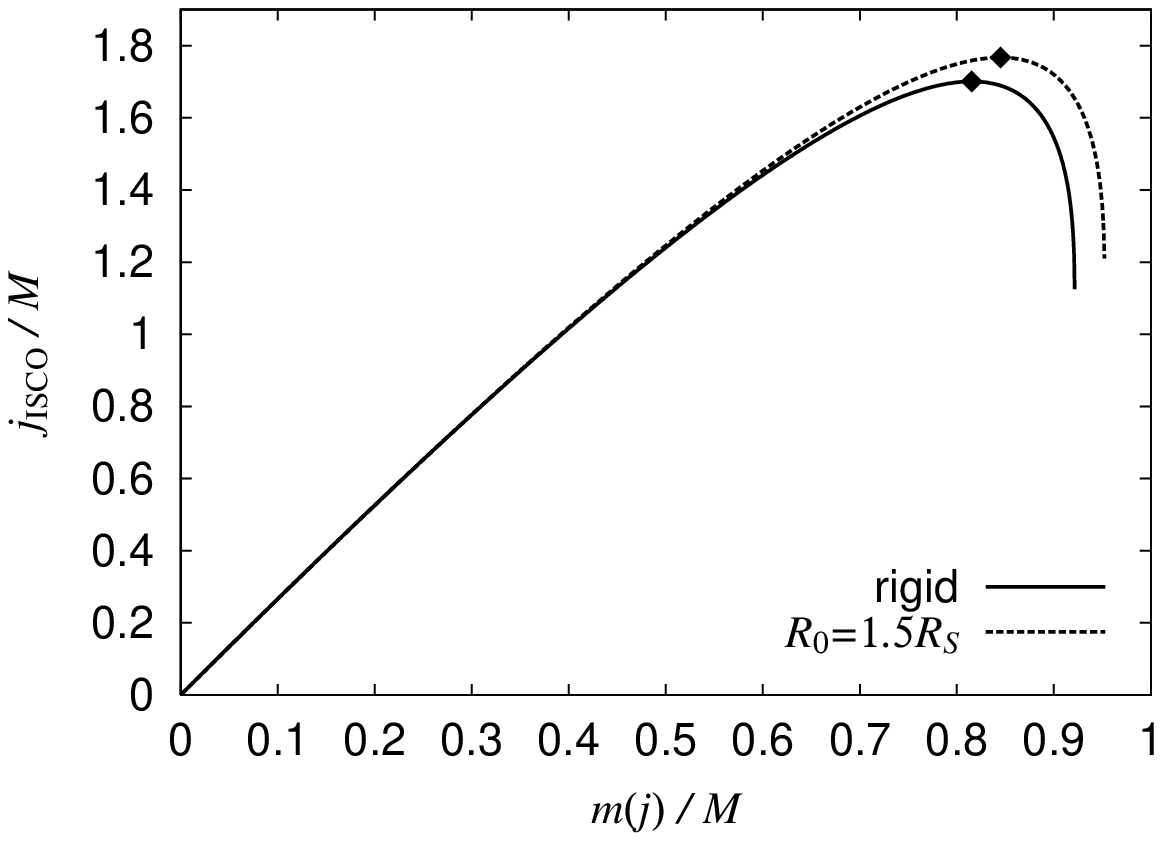}
\end{center}
\caption{(a)The distribution of $q(j)$ 
and (b) $j_{\rm ISCO}$ as a function of $m(j)/M(=m_{\ast}(j)/M_{\ast})$
for models with $\Omega_{0} = 0.065$ and for $\Gamma = 1.5$. 
The solid and dashed curves denote the results for
rigid rotation and differential rotation with $R_{0} = 1.5R_{S}$, 
respectively.
The filled diamonds denote values of $m(j)/M(=m_*(j)/M_*)$ and 
$q(j)$ at the maximum value of $j_{\rm ISCO}$. 
}\label{figure1}
\end{figure}
\begin{figure}[hctb]
\begin{center}
\epsfxsize=3.3in
\leavevmode
(a)\epsffile{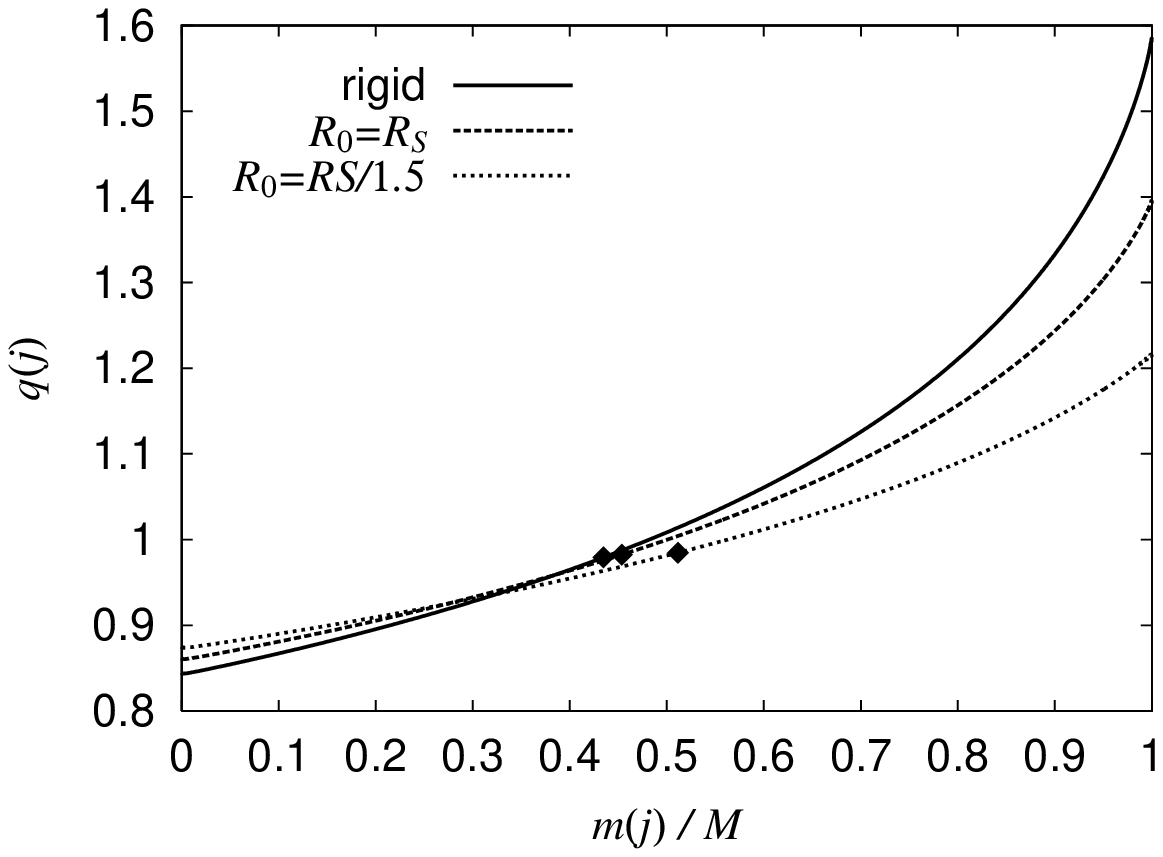}\\
\epsfxsize=3.3in
\leavevmode
(b)\epsffile{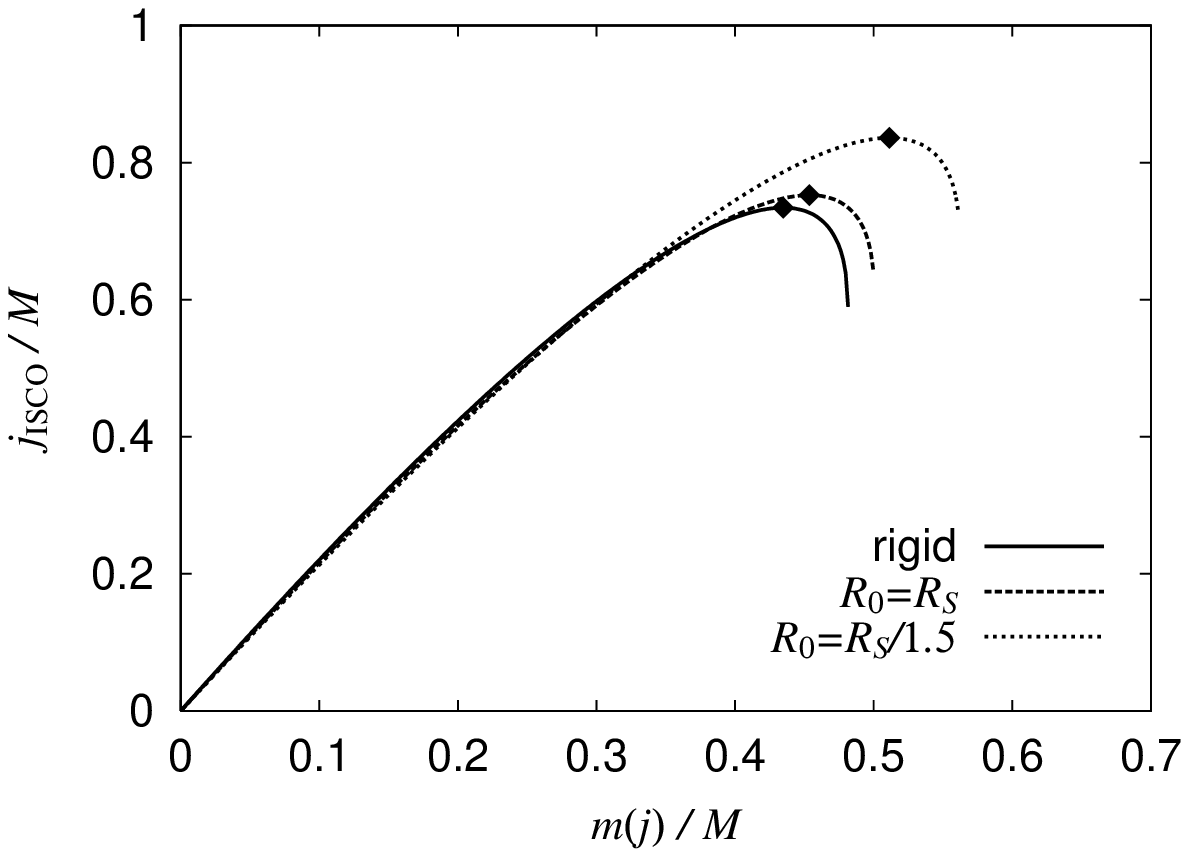}
\end{center}
\caption{The same as Fig. \ref{figure1} but 
for $\Omega_{0} = 0.090$. 
The solid, dashed, and dotted curves denote the results for
rigid rotation, differential rotation with $R_{0} = R_{S}$, and with
$R_{0} = R_{S}/1.5$, respectively. 
}\label{figure2}
\end{figure}
\begin{figure}[hctb]
\begin{center}
\epsfxsize=3.3in
\leavevmode
(a)\epsffile{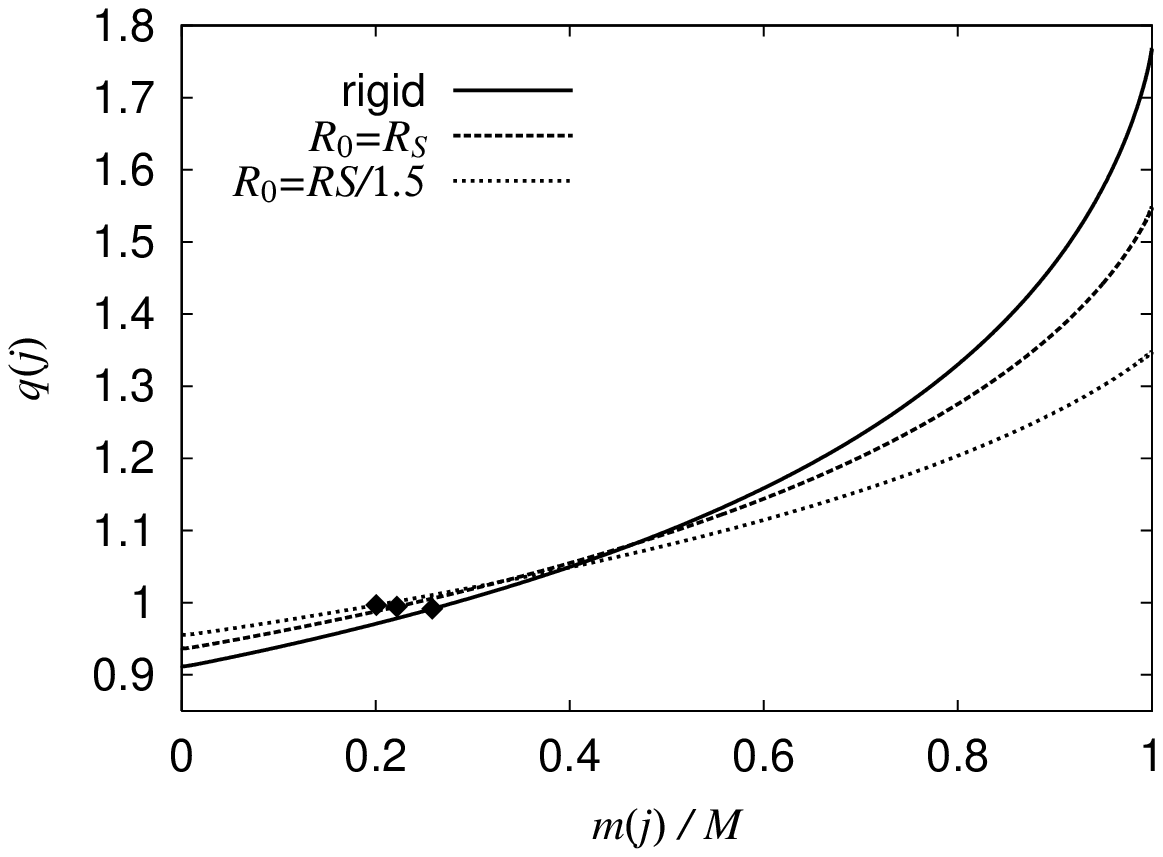}\\
\epsfxsize=3.3in
\leavevmode
(b)\epsffile{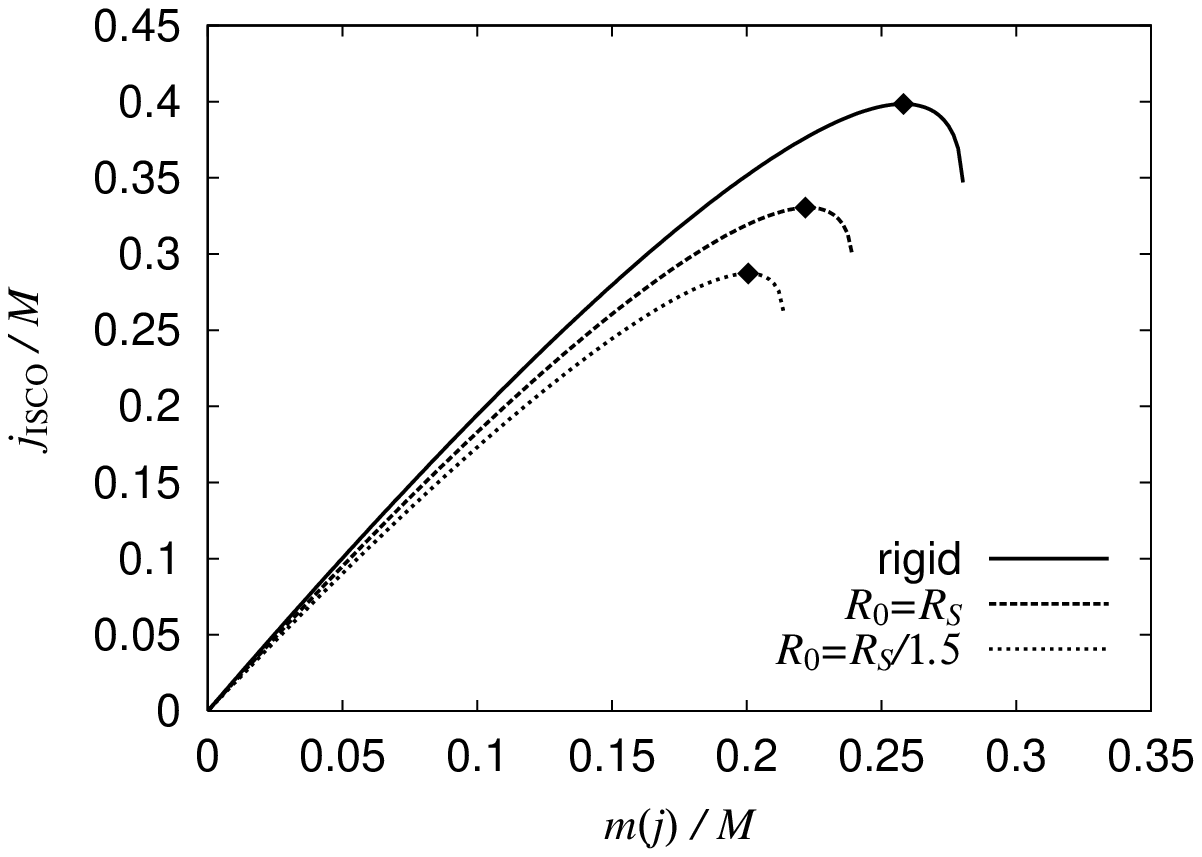}
\end{center}
\caption{The same as Fig. \ref{figure2} but 
for model of $\Omega_{0} = 0.100$. 
}\label{figure3} 
\end{figure}
\begin{figure}[hctb]
\begin{center}
\epsfxsize=3.3in
\leavevmode
\epsffile{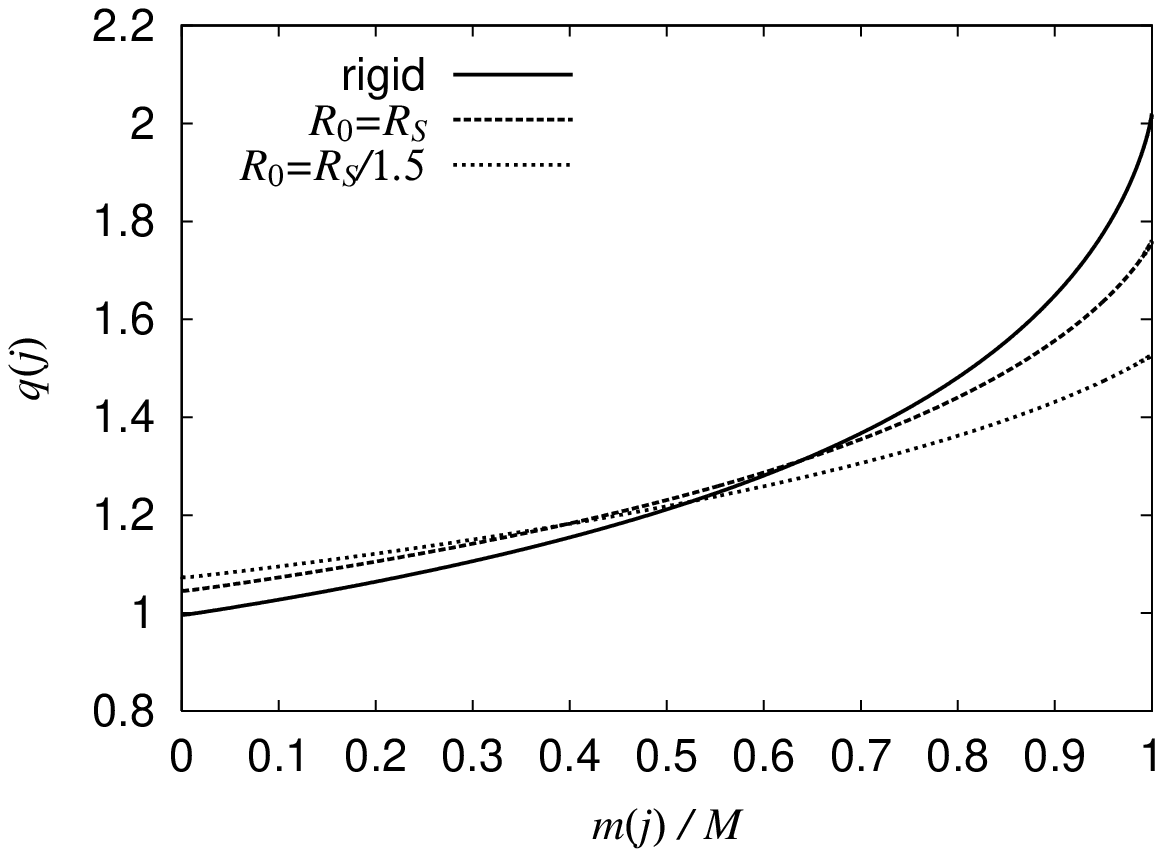}
\end{center}
\caption{The same as Fig. \ref{figure2}(a) but 
for model of $\Omega_{0} = 0.115$. 
}\label{figure4}
\end{figure}
\begin{figure}[hctb]
\begin{center}
\epsfxsize=3.3in
\leavevmode
\epsffile{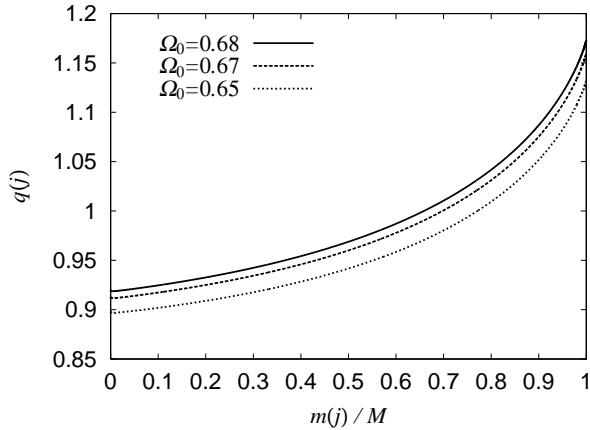}
\end{center}
\caption{The same as Fig. \ref{figure4} but for $\Gamma = 2.0$.
The solid, dashed, and dotted curves denote the results for
$\Omega_{0} = 0.68$, $0.67$, and $0.65$ respectively. All the models are
rigidly rotating cases.}\label{figure5}
\end{figure}
\begin{table}[htbc]
 \begin{center}
  \begin{tabular}{ccccc}
   $\Omega_{0}$ & $R_{0}$ & $\jisom/M$ & 
   $m_{\ast}(\jisom)/M_{\ast}$ & $q(\jisom)$ 
   \\ \hline 
   $0.050$ & $\infty$ (rigid) & 2.29 & 0.953 & 0.792  
   \\ \hline
   $0.065$ & $\infty$ (rigid) & 1.70 & 0.815 & 0.904 \\
   $0.065$ & $R_{S}/1.5$ & 1.77 & 0.845 & 0.902
   \\ \hline
   $0.090$ & $\infty$ (rigid) & 0.734 & 0.435 & 0.979  \\
   $0.090$ & $R_{S}$ & 0.753 & 0.454 & 0.982 \\
   $0.090$ & $1.5 R_{S}$ & 0.836 & 0.511 & 0.984 \\ 
   \hline
   $0.100$ & $\infty$ (rigid) & 0.399 & 0.256 & 0.992  \\
   $0.100$ & $R_{S}$     & 0.330 & 0.222 & 0.995 \\
   $0.100$ & $1.5 R_{S}$ & 0.287 & 0.201 & 0.997 \\ 
  \end{tabular}
 \end{center}
\caption{The maximum values of $j_{\rm ISCO}$, and corresponding values of 
 $m_{\ast}(j_{\rm ISCO})/M_{\ast}$ and $q(j_{\rm ISCO})$ for $\Gamma=1.5$ 
 initial models of $\Omega_{0} = 0.050$, 0.065, $0.090$, and $0.100$.
}\label{Table2}
\end{table}

Using the same method as described in \cite{ShibaSA}, 
we predict the final outcome of
the collapse for our initial conditions with $q > 1$. 
We pay particular attention to the black hole formation 
assuming that (i) the collapse proceeds in an axisymmetric manner, (ii)
the viscous angular momentum transport during the collapse is
negligible, and (iii) the pressure or heating effects never halt the
collapse. Because of the assumption that 
the viscous effect is negligible during the
collapse, the specific angular momentum $j$ of each fluid
element is conserved in an axisymmetric system. Here, $j$ is defined as
\beq
j \equiv h u_{\varphi}.
\eeq
Then we define a rest-mass distribution $m_{\ast}(j)$ as a function of $j$,
which is the integrated baryon rest mass of fluid elements with the
specific angular momentum less than $j$:
\beq
m_{\ast}(j) \equiv 2\pi \int _{j'< j} \rho_{\ast} r^{2}dr d(\cos \theta).
\eeq
Similarly, a specific angular momentum distribution is defined according to
\beq
J(j) \equiv 2\pi \int _{j'< j} \rho_{\ast} j' r^{2}dr d(\cos \theta).
\eeq
For the following analysis, it is better to have a quasi-local 
gravitational mass since in contrast with the case of soft equations
of state studied in \cite{ShibaSA}, 
the rest mass and the gravitational mass are different and
their difference is not negligible for models adopted in this paper 
(see Table I). However, the gravitational mass cannot be locally
defined in general relativity. Thus, here, 
assuming the ratio of the quasi-local gravitational mass 
to the rest mass is uniform inside a star, 
we define a ``gravitational mass distribution'' as 
\beq
m(j) \equiv {M \over M_*} m_*(j). 
\eeq
Note that $m(j)$ is equal to $M$ for a maximum value of $j$ (hereafter
$j_{\rm max}$) and that for $\Gamma \sim 4/3$, $m(j)$ is
approximately identical with $m_*(j)$. 

{}From these distribution functions, we define spin parameter
distributions as
\beq
q_{\ast}(j) \equiv \frac{J(j)}{m_{\ast}(j)^{2}} ~~~{\rm and}~~~
q(j) \equiv {J(j) \over m(j)^2}.  
\eeq
These may be approximately regarded as the non-dimensional spin 
parameters of fluid elements with the specific angular momentum
less than $j$. Although it is not clear which of two spin parameters
is better for the analysis, we adopt $q(j)$ 
because of the following reasons: (i) since 
$q(j)$ is equal to the global quantity 
$q$ for $j=j_{\rm max}$, $q(j)$ would be the better quantity for
a large value of $j$; 
(ii) $M_{\ast}$ is always larger than $M$ for all the models.
(This is likely to be the case for all the stable stars.) 
This implies that $q(j) > q_*(j)$. As a result, for $q(j) < 1$, 
$q_*(j)$ is always smaller than unity which is the key value for
black hole formation. 

In Figs. \ref{figure1}(a), \ref{figure2}(a), \ref{figure3}(a),
and \ref{figure4}, $q(j)$ as a function of $m_*(j)/M_*$
are displayed for the parameters listed in Table I for $\Gamma = 1.5$. 
Figure \ref{figure5} also shows the same relation for parameters listed
in Table II for $\Gamma = 2.0$. 
These figures indicate that the value of $q(j)$ 
at the center of stars (hereafter denoted as $q_{c}$ and $q_{*,c}$) 
can be much smaller than unity 
even if the global values of $q$ and $q_*(j_{\rm max})$
(hereafter denoted simply as $q_{\ast}$) are larger than unity. 
An outstanding difference between the results for
two values of $\Gamma$ is 
the ratio of $q/q_c$ for the rigidly rotating case,
which is $\sim 2$ for $\Gamma=1.5$ while $\sim 1.25$ for $\Gamma=2$.
This results from the fact that for softer equations of state, the
star has a more centrally-condensed structure. 

As collapse proceeds, an inner region of the stars collapses faster. This
property will be more outstanding for the softer equations of state, with which
stars have a more centrally-condensed structure.
Taking this and the distribution of $q(j)$  into account, 
we conjecture that an inner region of $q(j) < 1$ 
may form a seed black hole first and the black hole
subsequently grows as the ambient fluids accrete onto it. 

Assuming that a seed black hole is formed during the collapse for models 
of $q_{c} < 1$, we predict the subsequent evolution in the following
manner. Here, we focus only on the rapidly rotating case with a
large value of $q>1$ but with $q_c <1$. Since 
the centrifugal force of the rapidly rotating progenitors is 
large, first, the collapse is likely to proceed along the rotational axis 
to be a disk structure. In such a disk, fluid elements of a cylindrical
radius are likely to have the approximately identical
value of the specific angular momentum $j$. 
As a result, the fluid elements of the same value of $j$ will collapse 
in a simultaneous manner even if they are initially at
different locations. Now, let us 
consider the innermost stable circular orbit (ISCO) around the growing
black hole at the center. If the value of $j$ of a fluid 
element is smaller than that of the seed black hole 
at the ISCO, $j_{\rm ISCO}$, the element 
will fall into the seed black hole eventually. In fact, there is a
possibility that some fluids can be captured even for $j > j_{\rm ISCO}$
if it is on a noncircular orbit. Ignoring such trajectories yields the
minimum amount of fluids that will fall into the black hole. The value of 
$j_{\rm ISCO}$ will change as ambient fluids accrete onto the black
hole. If $j_{\rm ISCO}$ increases as a result of the accretion, 
the more ambient fluids will fall into the black hole. 
This suggests that the evolution of mass $m_{\ast}(j)$ and angular momentum
$J(j)$ inside the seed black hole will approximately proceed 
according to the initial distribution of $m_{\ast}(j)$ and $J(j)$
because all viscous effects such as angular momentum transfer
are assumed to be negligible. 
On the other hand, if $j_{\rm ISCO}$ decreases during the accretion, no 
more fluid will fall into the black hole, and as a result, the
dynamical growth of the black hole will terminate.

To estimate the value of $j_{\rm ISCO}$ and to predict the 
growth path of the seed black hole, we assume that the spacetime metric can be
instantaneously approximated by that of a Kerr spacetime of mass 
$m(j)$ and spin $q(j)$.
On these approximations, we can compute $j_{\rm ISCO}$ of a 
seed black hole as \cite{ST,BPT,FN}, 
\beq \label{def-jISCO}
 j_{\rm ISCO} = 
\frac{ \sqrt{m(j) \riso}
\left( \riso^{2} - 2 q(j)m(j) \sqrt{m(j) \riso} 
+ (q(j)m(j))^{2}\right)}
{\riso \left(\riso^{2} -3m(j) \riso 
+ 2 q(j)m(j) \sqrt{m(j) \riso}\right)^{1/2}} , 
\eeq
where
\beqn
&& \riso = m(j) \left[ 3 + Z_{2} -
	 \left\{(3-Z_{1})(3+Z_{1}+2Z_{2})\right\}^{1/2} \right],
\nonumber \\
&& \ \ Z_{1} = 1 
+ \left[1- q(j)^{2} \right]^{1/3}
\left[ \{1+ q(j)\}^{1/3} + \{1-q(j)\}^{1/3}\right], \nonumber \\
&& \ \ Z_{2} = \left[ 3q(j)^{2} + Z_{1}^{2} \right]^{1/2} . \nonumber 
\eeqn
Here, $m(j)$ and $q(j)$ are not strictly
the gravitational mass and the spin parameter. 
We suspect that they may have a systematic error of magnitude
$\alt |1-M/M_*|$ for the mass and $\alt 2|1-M/M_*|$ for the
spin parameter. This implies that the estimation
of $j_{\rm ISCO}$ by Eq. (\ref{def-jISCO}) also includes a systematic
error. The magnitude of the error for $j_{\rm ISCO}$ may be $\alt 10$\%. 

In Figs. \ref{figure1}(b), \ref{figure2}(b), and \ref{figure3}(b), 
we show $j_{\rm ISCO}[ m(j), q(j) ]$ as
a function of $m_{\ast}(j)/M_{\ast}$ for $\Gamma=1.5$. These figures
show that for models in which the degree of differential rotation is
not too high, $j_{\rm ISCO}$ has a maximum (hereafter denoted as 
$\jisom$). 
Thus, we predict that the seed black hole will grow until $j$
reaches $\jisom$. 
In Table \ref{Table2}, we show $\jisom$, 
$m_{\ast}(\jisom)$, and $q(\jisom)$ for 
$\Gamma=1.5$. Note that for models of $\Omega_{0} = 0.115$ in
which $q_{c} > 1$, we predict that a seed black hole is not formed. 
Now, let us assume that $m_{\ast}(\jisom)/M_{\ast}$ and 
$q(\jisom)$ may be approximately regarded as the
mass fraction $M_{\rm BH}/M$ and spin parameter
$q_{\rm BH}$ of the final black hole (which is in a quasi-stationary
state) as 
\beqn
&&M_{\rm BH} \approx m(\jisom), 
\label{approx-MBH} \\
&&q_{\rm BH} \approx q(\jisom)
\label{approx-JBH}.
\eeqn
{}From this approximation we immediately predict
that an appreciably massive disk 
($M_{\rm disk}/M = 0.1$--0.8) will be formed after stellar collapse. 
Note that $M_{\rm BH}$ and $q_{\rm BH}$ computed above 
might include a systematic error because of the reason that we 
approximate a spacetime composed of a 
black hole and a massive disk simply as a Kerr spacetime. However, 
$j_{\rm ISCO}$ is likely to be determined from the local quantities near
the black hole and, hence, we consider that our treatment may
be approximately correct.

It should be addressed that for a fixed value of $q_{c}$, the
corresponding value of $q(\jisom)$ is almost 
independent of the differential rotation parameter $R_{0}$,
i.e., the value of $q$.
This suggests that the spin parameter of the 
final state of a black hole may be determined by the
value of $q_{c}$ and independent of $q$ as far as $q(j)$ is
an increasing function of $j$. 

In summary, we have predicted the following facts in this section:
(I) inner region of a star with $q_{c} < 1$ will collapse first to form
a seed black hole even if the global value $q$
exceeds unity significantly; 
(II) the formed seed black hole
will grow as the ambient fluids subsequently accrete onto it, 
(III) evolution of the relation between 
the rest mass $m_{\ast}(j)$ and angular momentum $J(j)$ enclosed
inside a growing black hole will agree approximately with
the initial relation between $m_{\ast}(j)$ and $J(j)$; 
(IV) whether black hole is formed or not will be 
determined by the value of $q_{c}$ independent of the value of $q$;  
(V) the final outcome of a dynamical 
collapse for a star with $q_{c}<1$ is a black hole surrounded by a disk.  

These predictions made form the initial conditions 
are quite reasonable, but to confirm them, 
it is obviously necessary to perform fully 
general relativistic simulations. The next section presents 
the results of numerical simulation and demonstrates that 
the predictions are correct. 

\section{RESULTS of NUMERICAL SIMULATION}
\subsection{Results for $\Gamma = 1.5$}

\begin{figure}[tb]
\begin{center}
\epsfxsize=3.3in
\leavevmode
(a)\epsffile{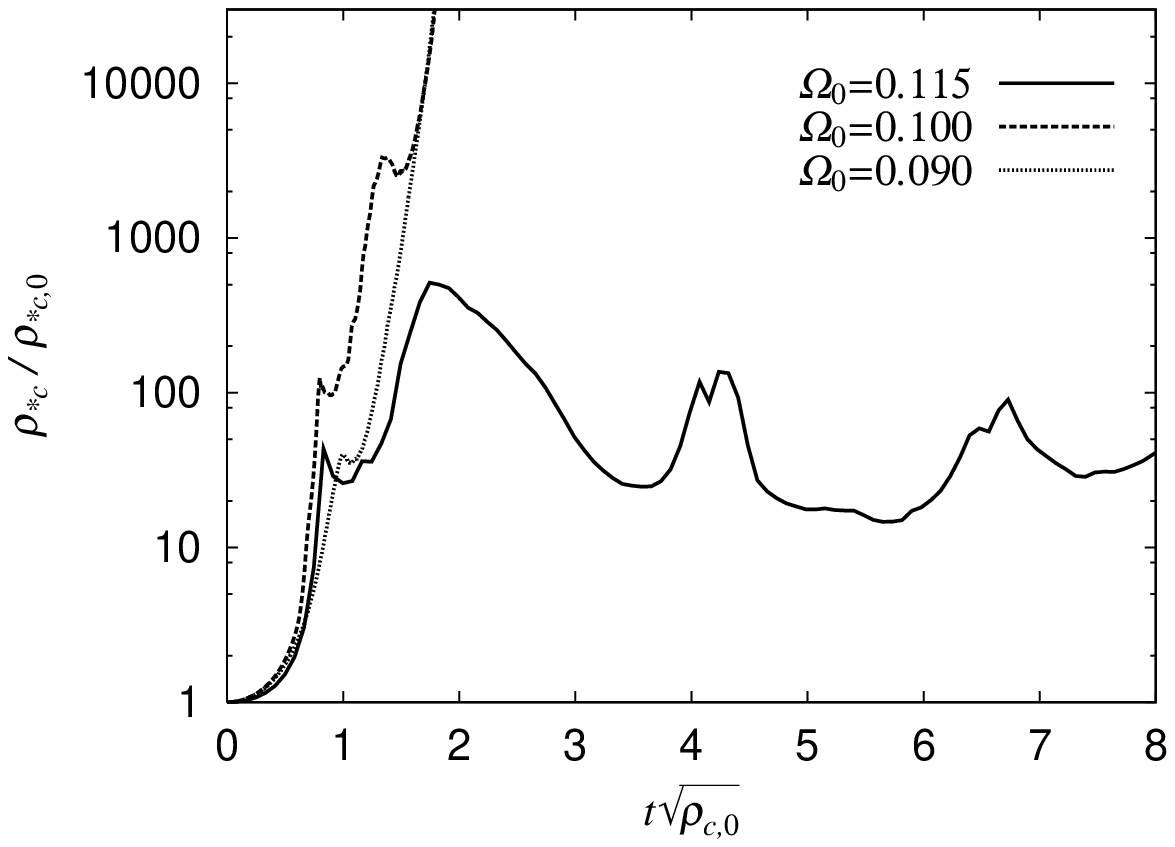}
\epsfxsize=3.3in
\leavevmode
(b)\epsffile{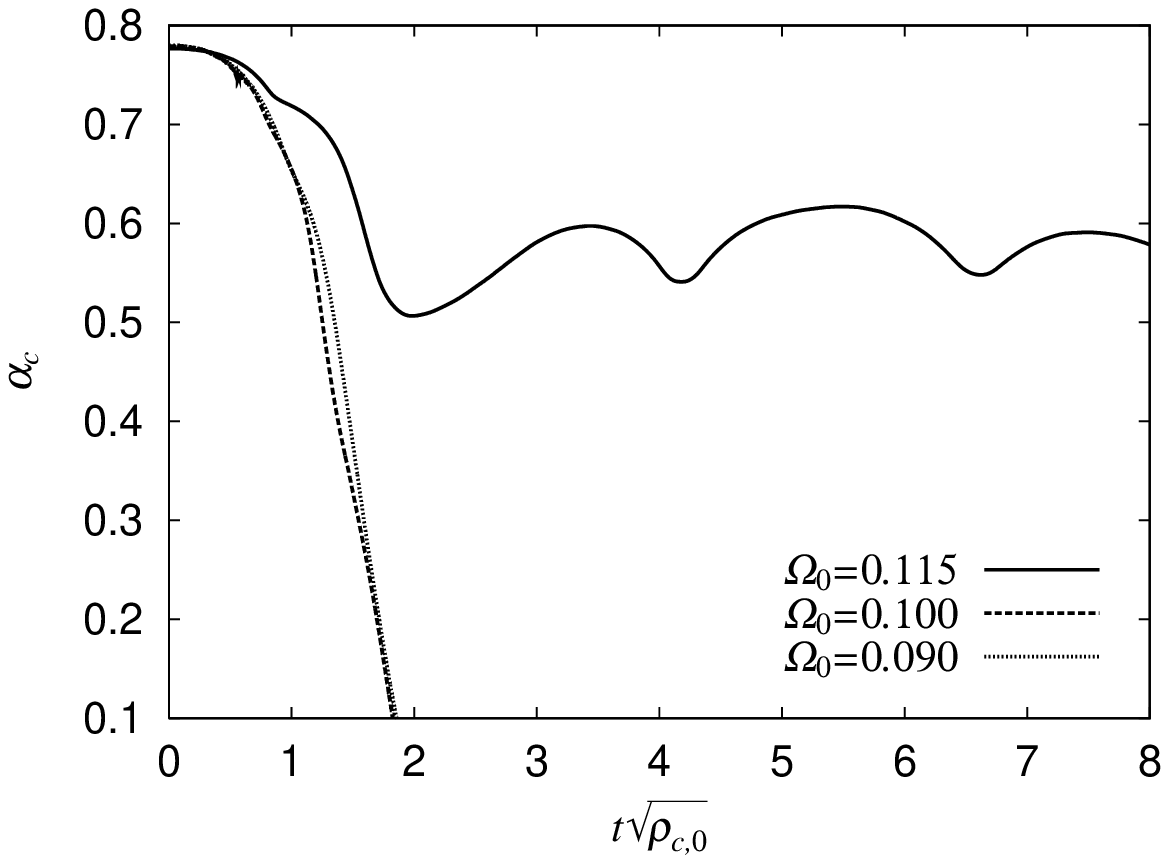}
\end{center}
\caption{Evolution of (a) the central density $\rho_{*,c}$ and (b) the
central value of the lapse function $\alpha_{c}$ for $\Gamma=1.5$. 
In both figures, solid, dashed, and dotted curves denote the 
differentially rotation models of $R_S=R_0/1.5$ with 
$\Omega_{0} = 0.090$, $0.100$, and $0.115$ 
($q_c=0.87$, 0.96, and 1.08), respectively. 
}\label{figure6}
\end{figure}
\begin{figure}[tb]
\begin{center}
\epsfxsize=2.25in
\leavevmode
\epsffile{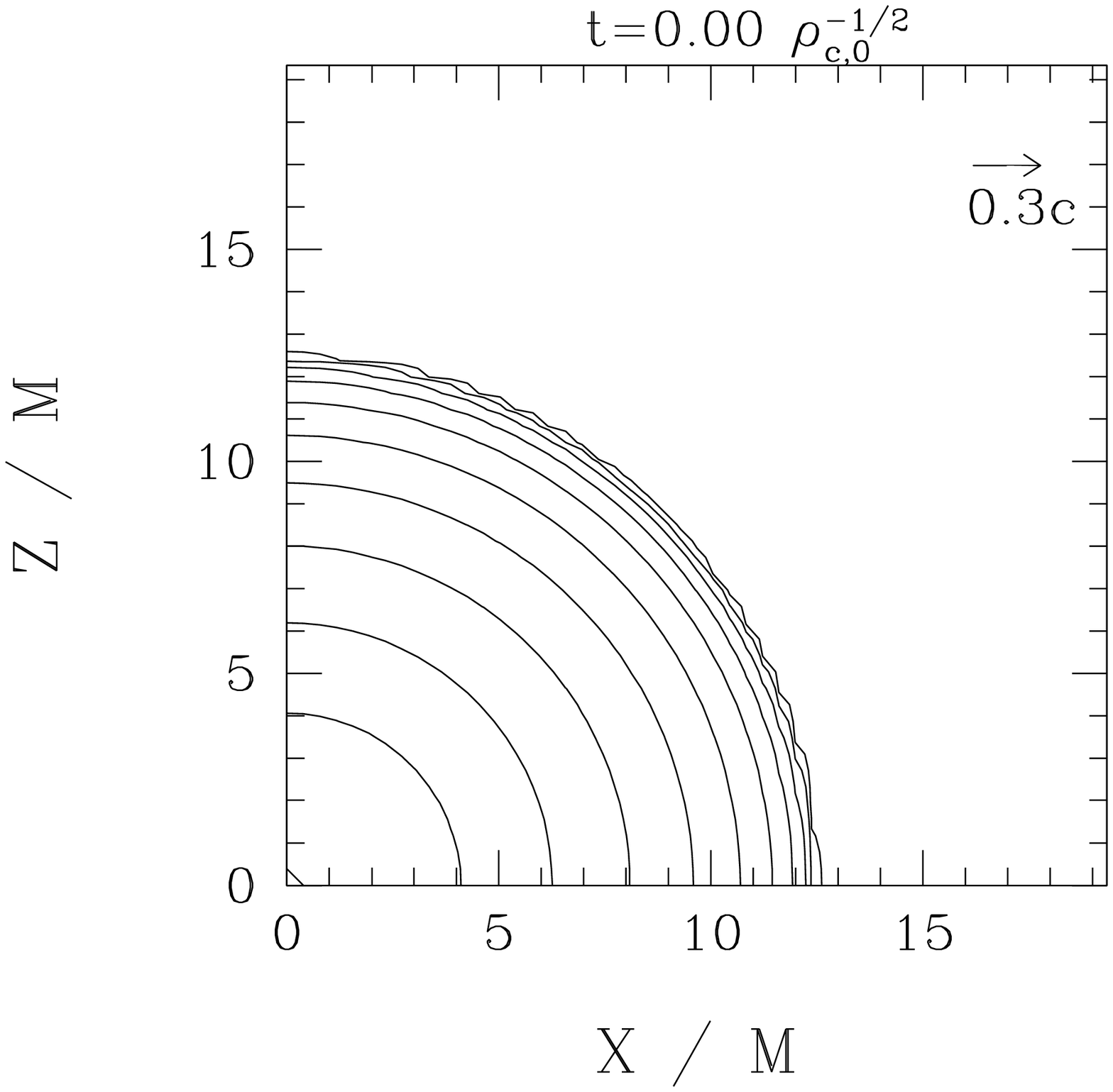}
\epsfxsize=2.25in
\leavevmode
\epsffile{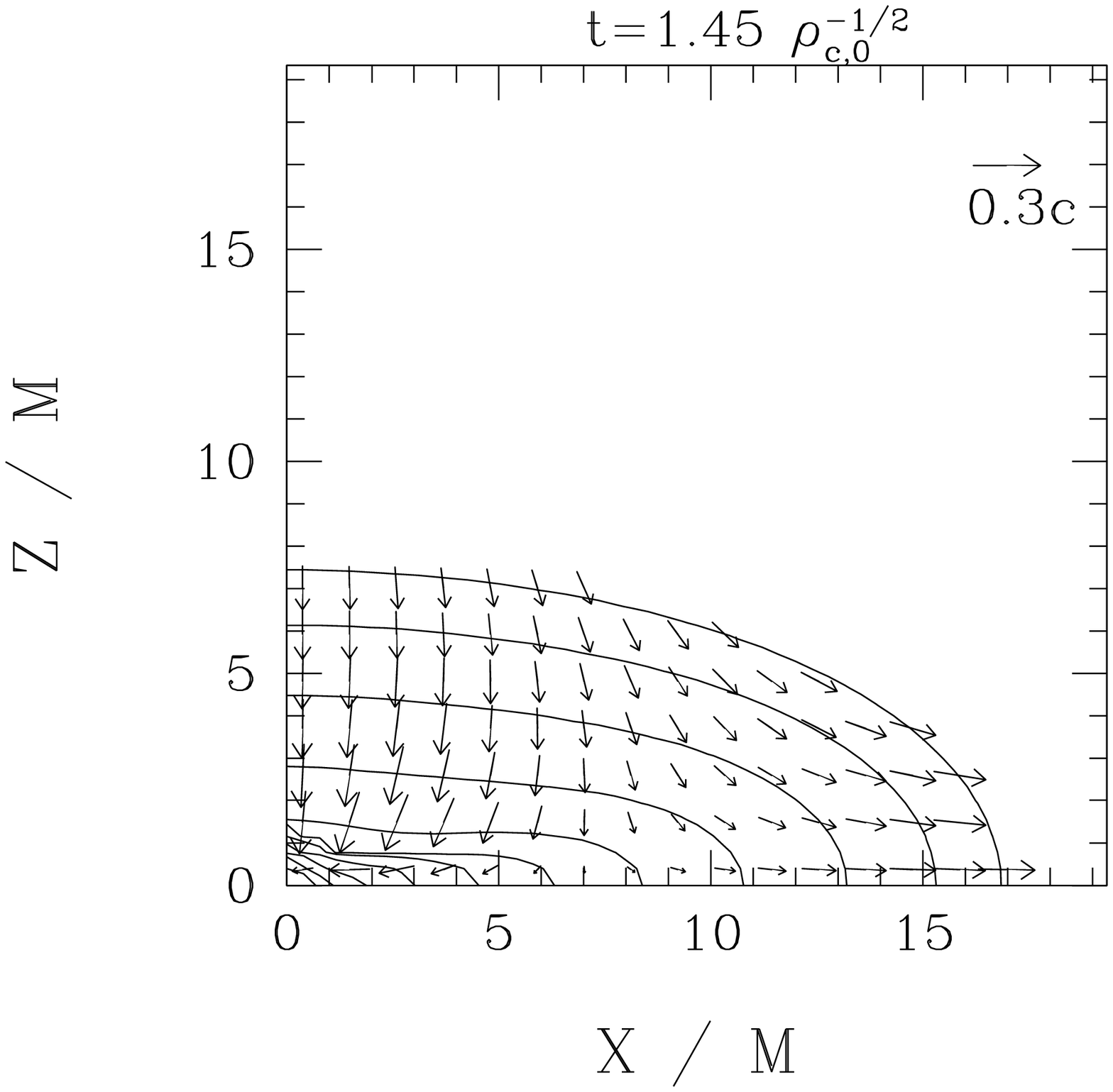}
\epsfxsize=2.25in
\leavevmode
\epsffile{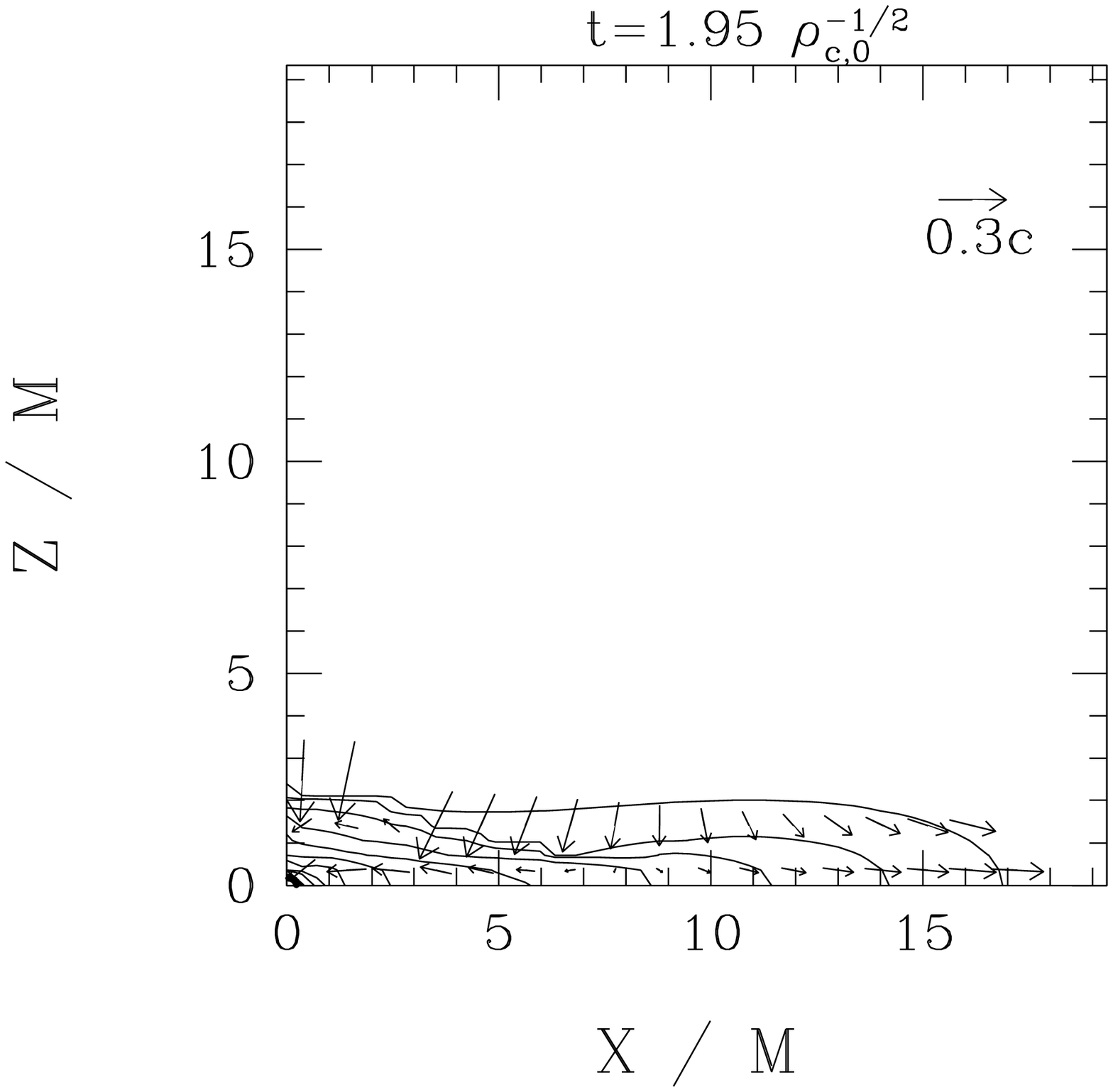} \\
\epsfxsize=2.25in
\leavevmode
\end{center}
\caption{Snapshots of the density contour curves of $\rho_*$ and
of the velocity field of $(v^x, v^z)$ at selected time slices
for the rigid rotation model with $\Omega_0=0.065$, 
The contour curves are drawn for $\rho_*=\rho_a \times 10^{-0.5j}$, 
with $j=0,1,2,\cdots,10$ where $\rho_{a}=0.011$, 1, and
10 at $t\rho_{c,0}^{1/2}=0$, 1.45, and 1.95. 
The thick solid circle 
with the radius $\sim 0.2 M$ of the last panel  
denotes the apparent horizon. 
}\label{figure7} 
\end{figure}
\begin{figure}[tb]
\begin{center}
\epsfxsize=2.25in
\leavevmode
\epsffile{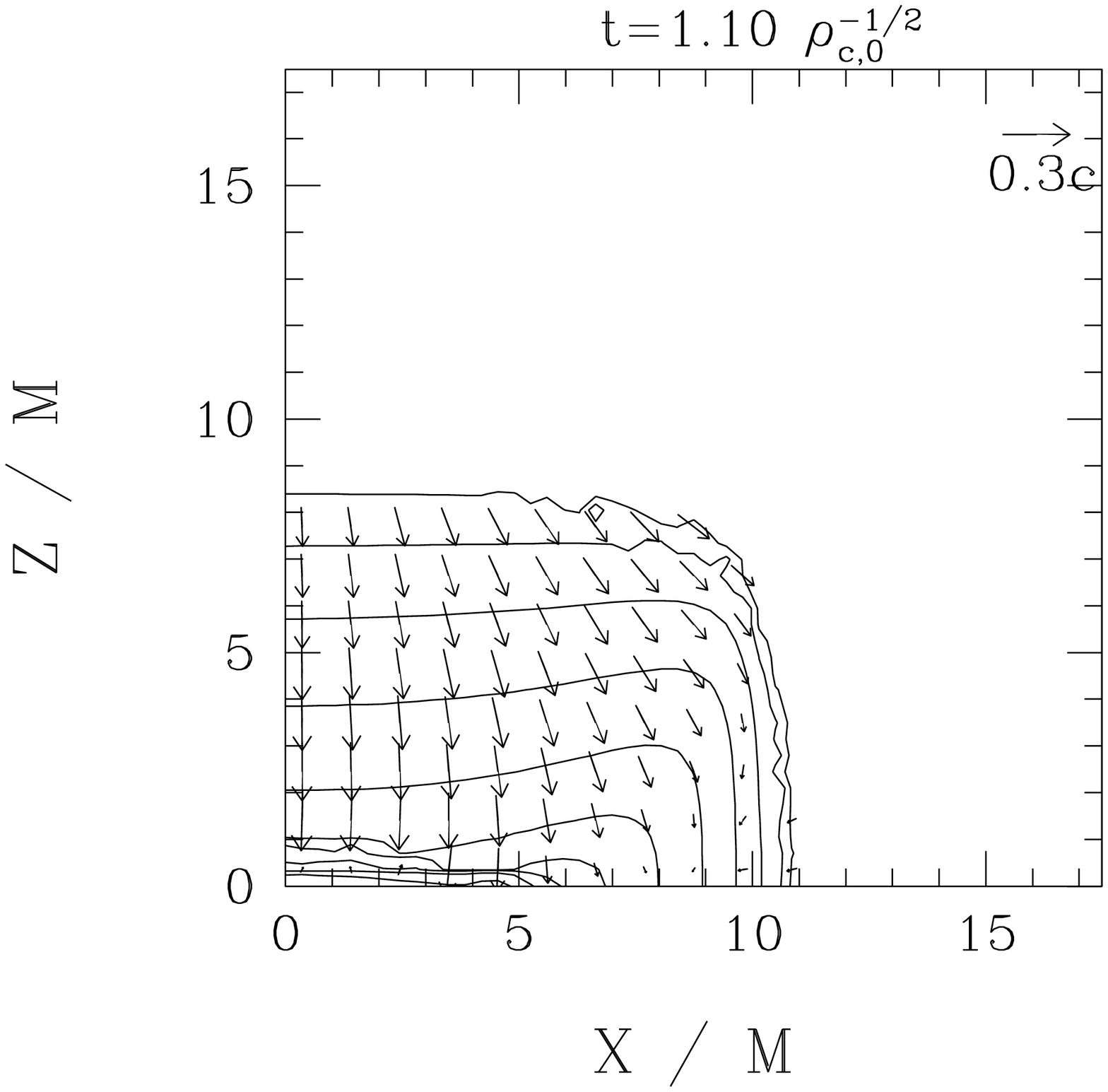}
\epsfxsize=2.25in
\leavevmode
\epsffile{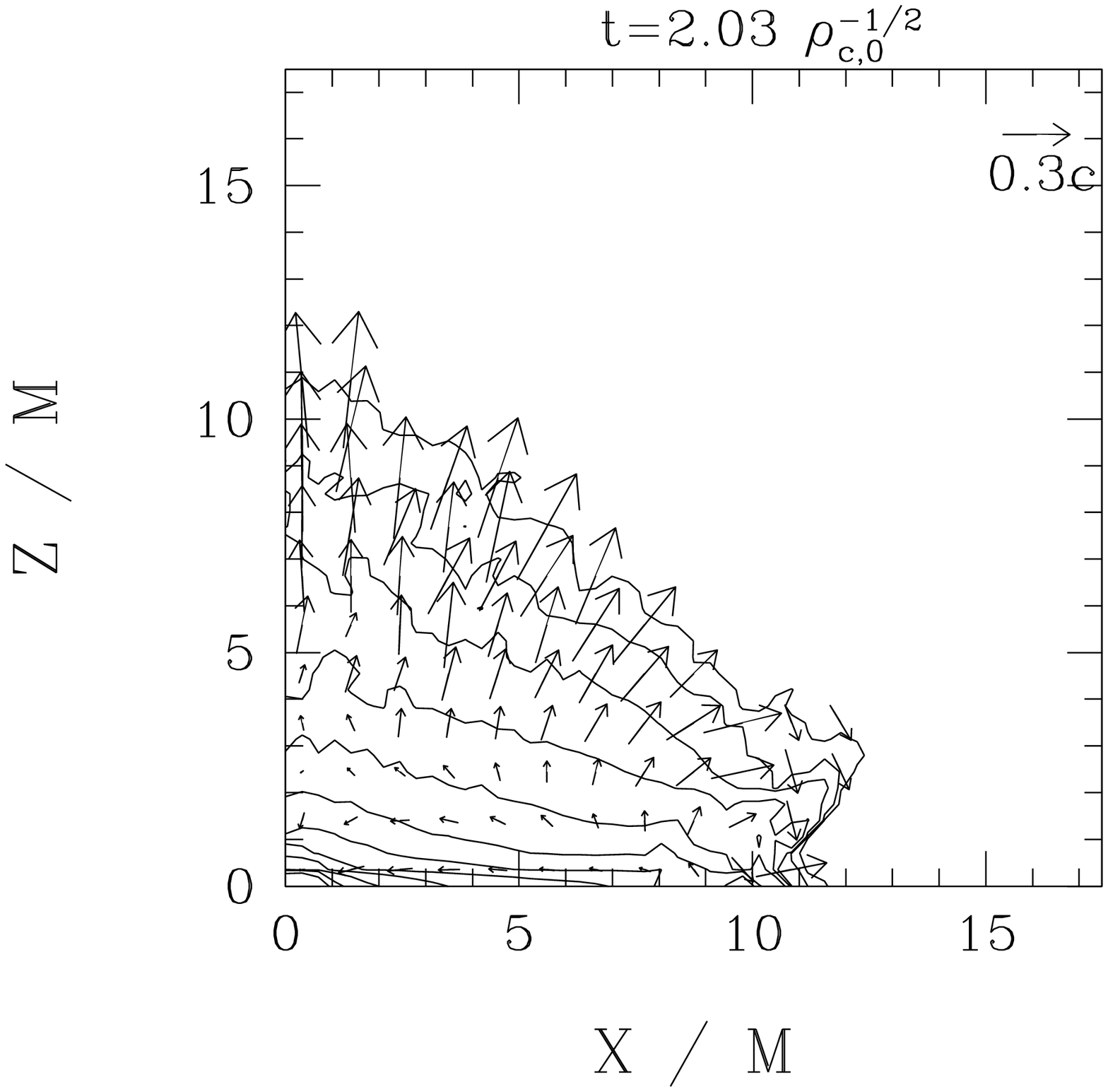}
\epsfxsize=2.25in
\leavevmode
\epsffile{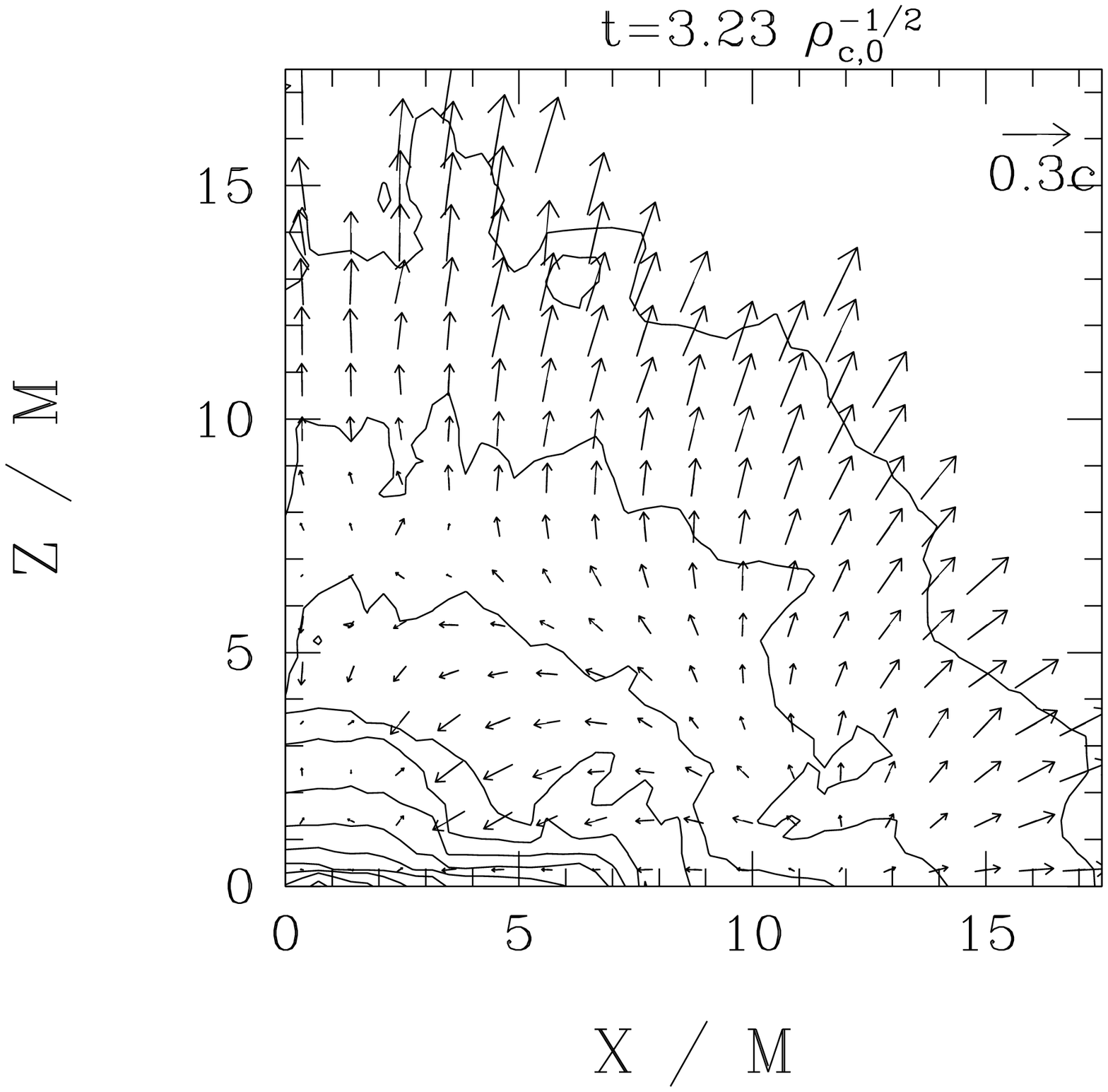}
\end{center}
\caption{The same as Figs. \ref{figure7} but 
for the differential rotation model with $\Omega_0=0.115$
and $R_0=R_S/1.5$. The contour curves are drawn for
$\rho_*=\rho_a \times 10^{-0.5j}$, 
with $j=0,1,2,\cdots,10$ where 
$\rho_a=1$ for all the time steps. 
}\label{figure8} 
\end{figure}
\begin{figure}[tb]
\begin{center}
\epsfxsize=3.3in
\leavevmode
(a)\epsffile{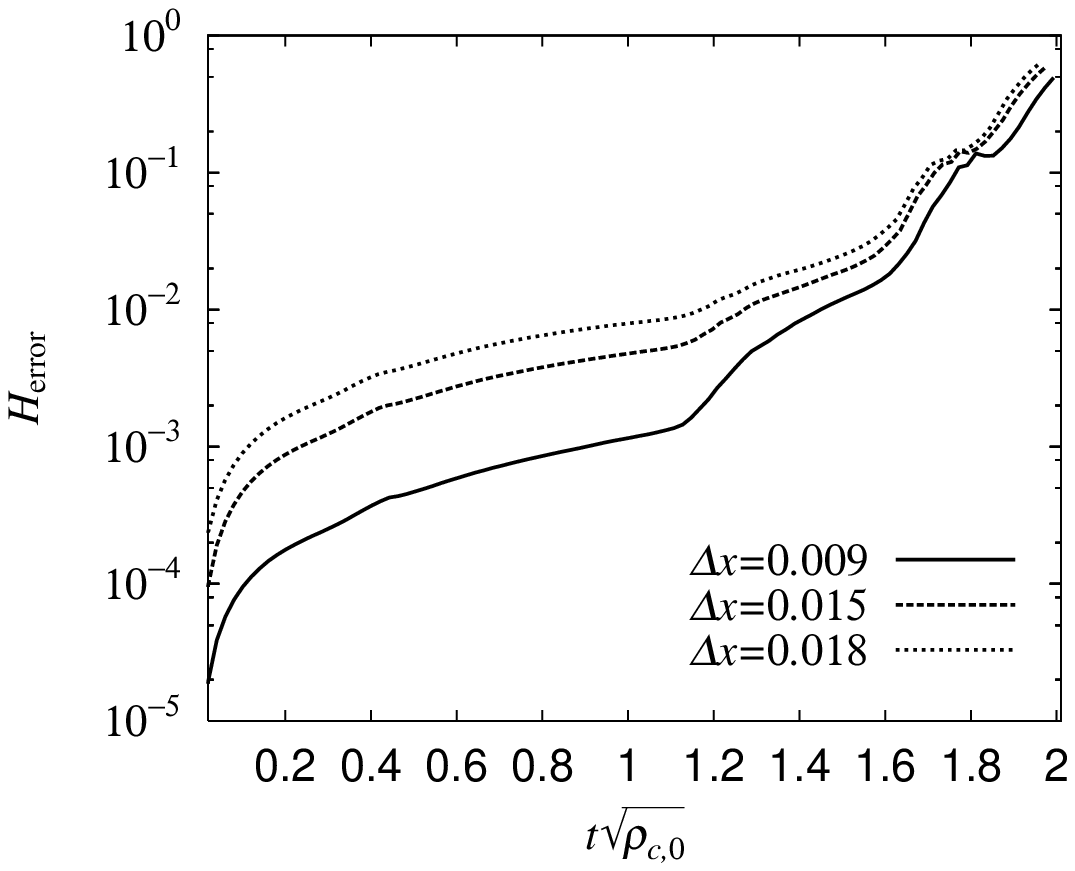}
\epsfxsize=3.3in
\leavevmode
(b)\epsffile{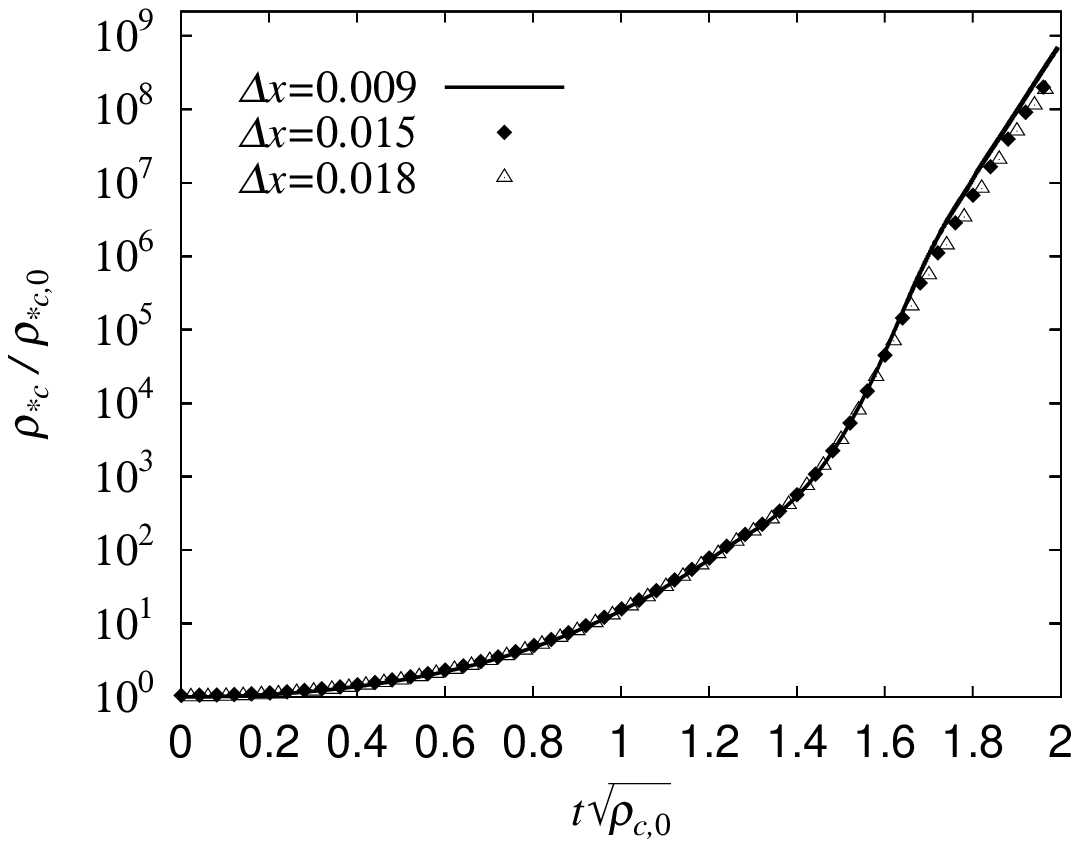}
\end{center}
\caption{Time evolution of (a) violation of the Hamiltonian constraint
 $H_{\rm error}$ and (b) the central value of $\rho_*$ ($\rho_{*,c}$)
for $\Omega_{0} = 0.065$ and $R_{0} = 1.5 R_{S}$. 
In panel (a), the solid, dashed, and dotted curves denote the results of 
$\Delta x = 0.009$, $0.015$, and $0.018$. In panel (b), the filled diamonds
and open triangles denote the results of 
$\Delta x = 0.015$ and $0.018$.  These figures indicate that 
second-order convergence is achieved in our simulations.
}\label{figure9} 
\end{figure}
\begin{figure}[hbtc]
\begin{center}
\epsfxsize=3.3in
\leavevmode
(a)\epsffile{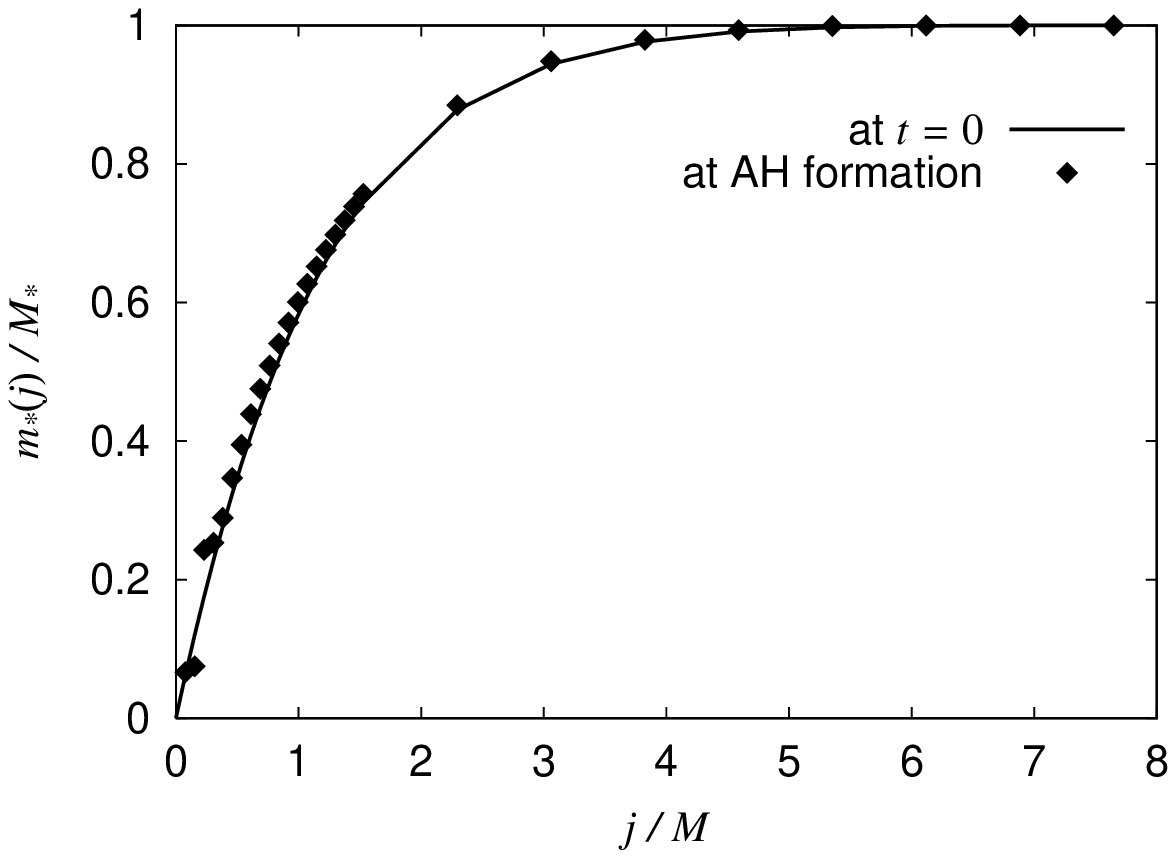}
\epsfxsize=3.3in
\leavevmode
(b)\epsffile{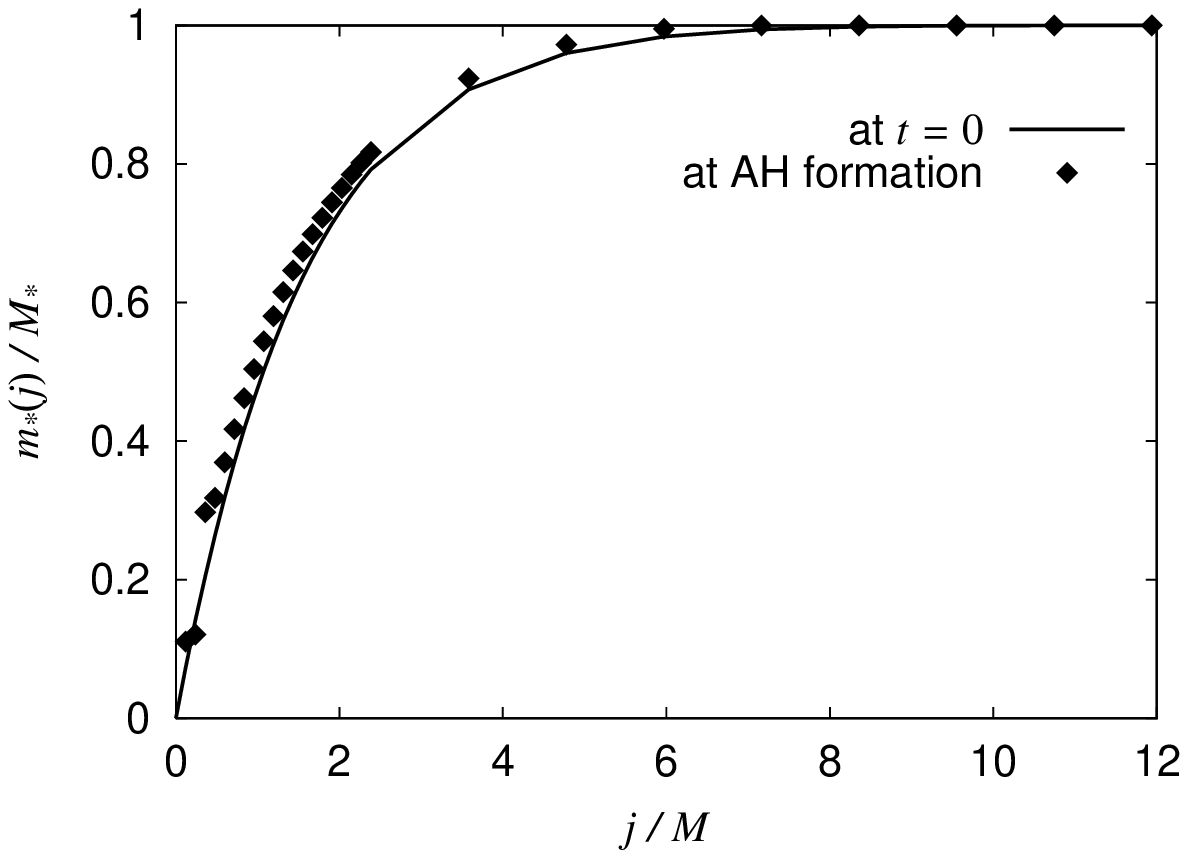}
\end{center}
\caption{Rest-mass distribution as a function of specific
 angular momentum $j$ (a) for the rigid rotation models with 
$\Omega_{0} = 0.065$ and (b) for $\Omega_{0} = 0.090$.
In all figures, the
solid curves denote the initial rest-mass
distributions and the filled diamonds
denote the rest-mass distributions when an apparent horizon is
first formed. The grid resolution is $\Delta x = 0.01125$ for (a) 
and $\Delta x = 0.006$ for (b).}\label{figure10} 
\end{figure}
\begin{figure}[htbc]
\begin{center}
\epsfxsize=3.3in
\leavevmode
(a)\epsffile{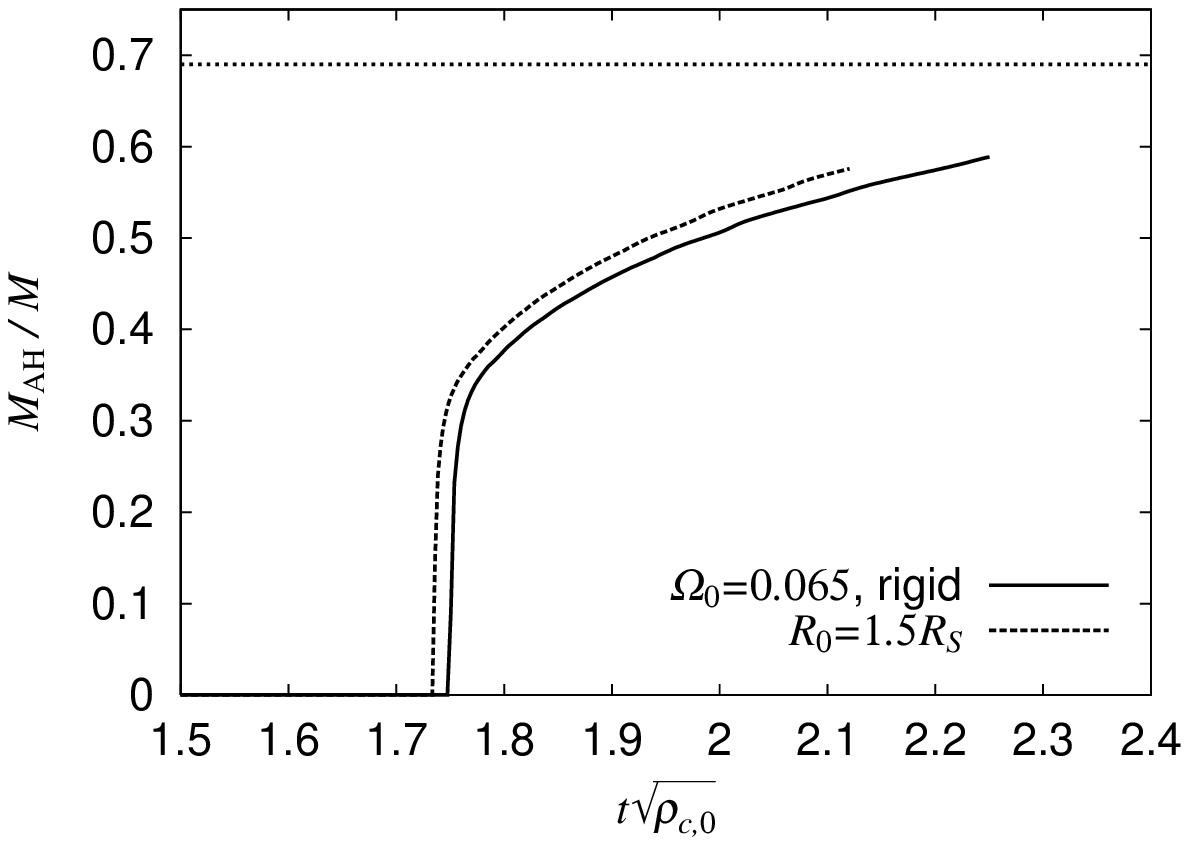}
\epsfxsize=3.3in
\leavevmode
(b)\epsffile{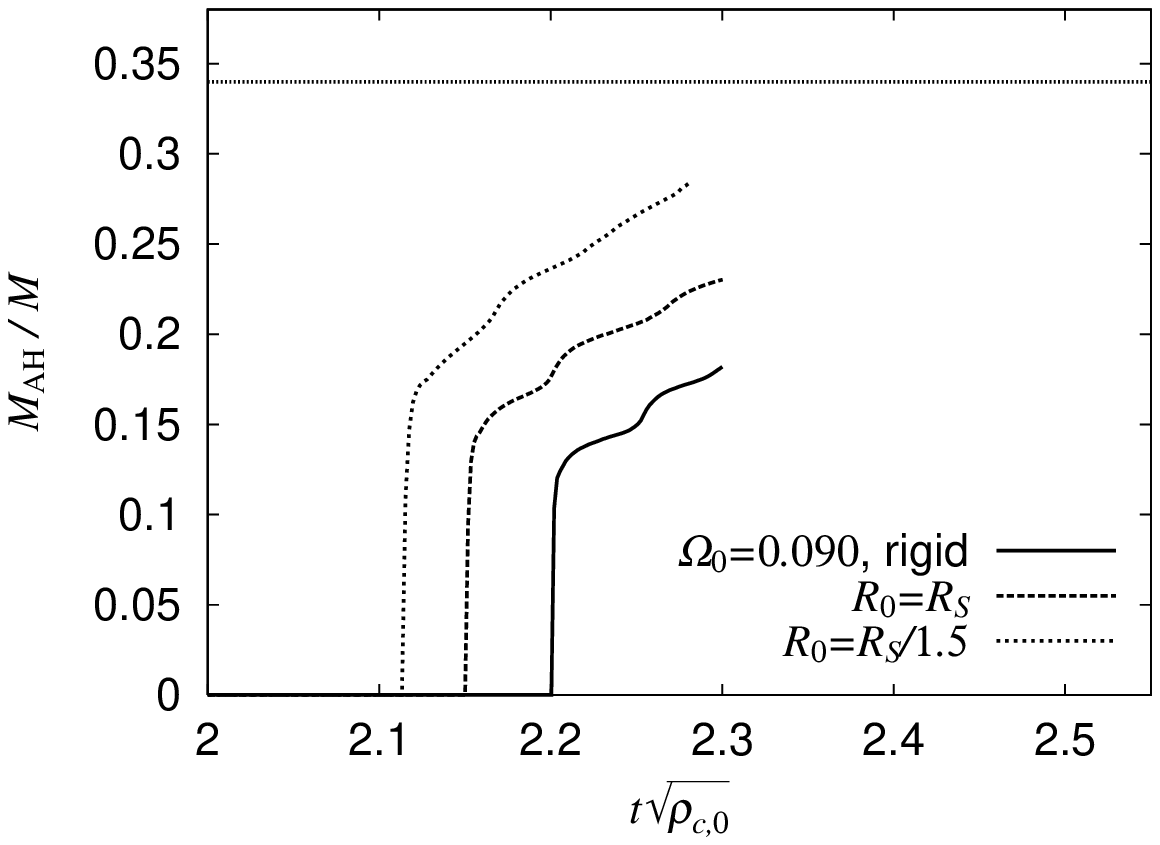}
\end{center}
\caption{Evolution of the apparent horizon mass $M_{\rm AH}$
for the rigid rotation models with 
(a) $\Omega_{0} = 0.065$ and (b) $\Omega_{0} = 0.090$.
In both figures the horizontal lines denote the irreducible mass
$M_{\rm irr}$ with the rigid rotation models,
$\approx 0.67M$ for $\Omega_{0} = 0.065$ and
$\approx 0.34M$ for $\Omega_{0} = 0.090$. 
}\label{figure11}
\end{figure}
\begin{figure}[htbc]
\begin{center}
\epsfxsize=3.3in
\leavevmode
(a)\epsffile{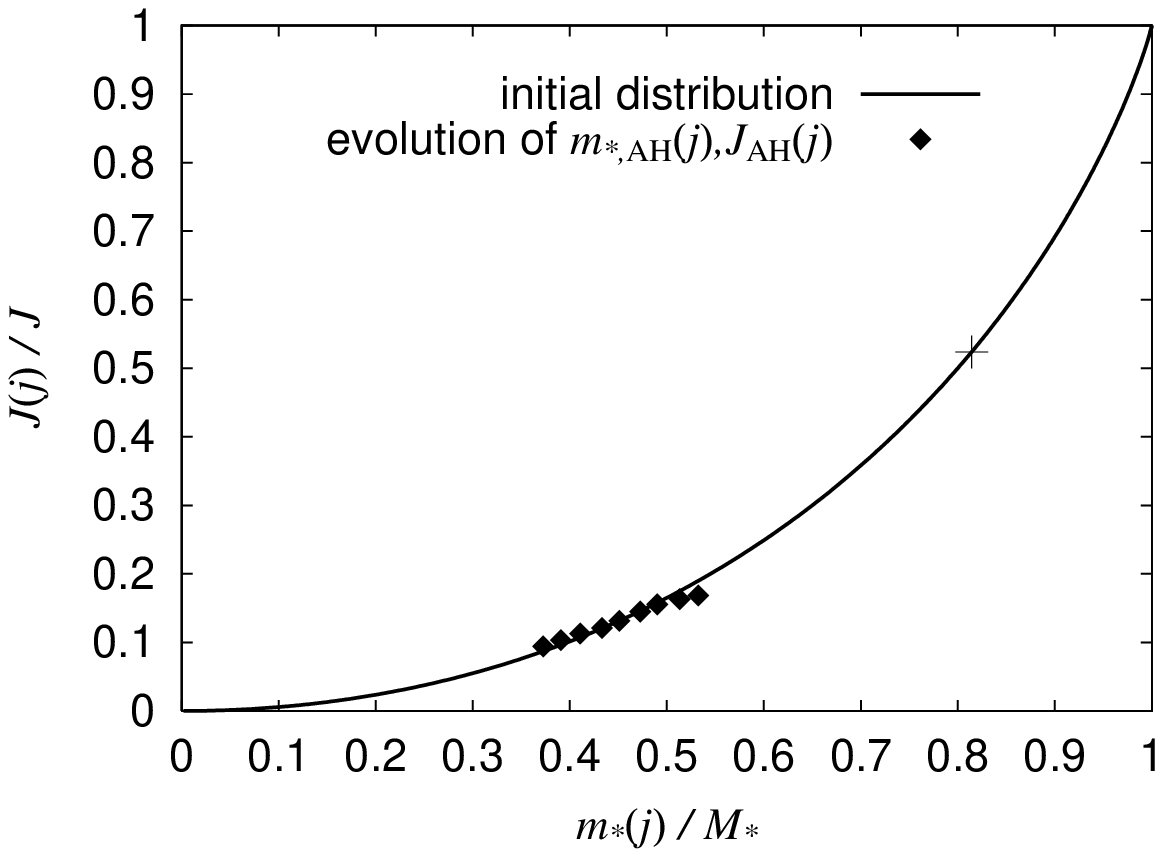}
\epsfxsize=3.3in
\leavevmode
(b)\epsffile{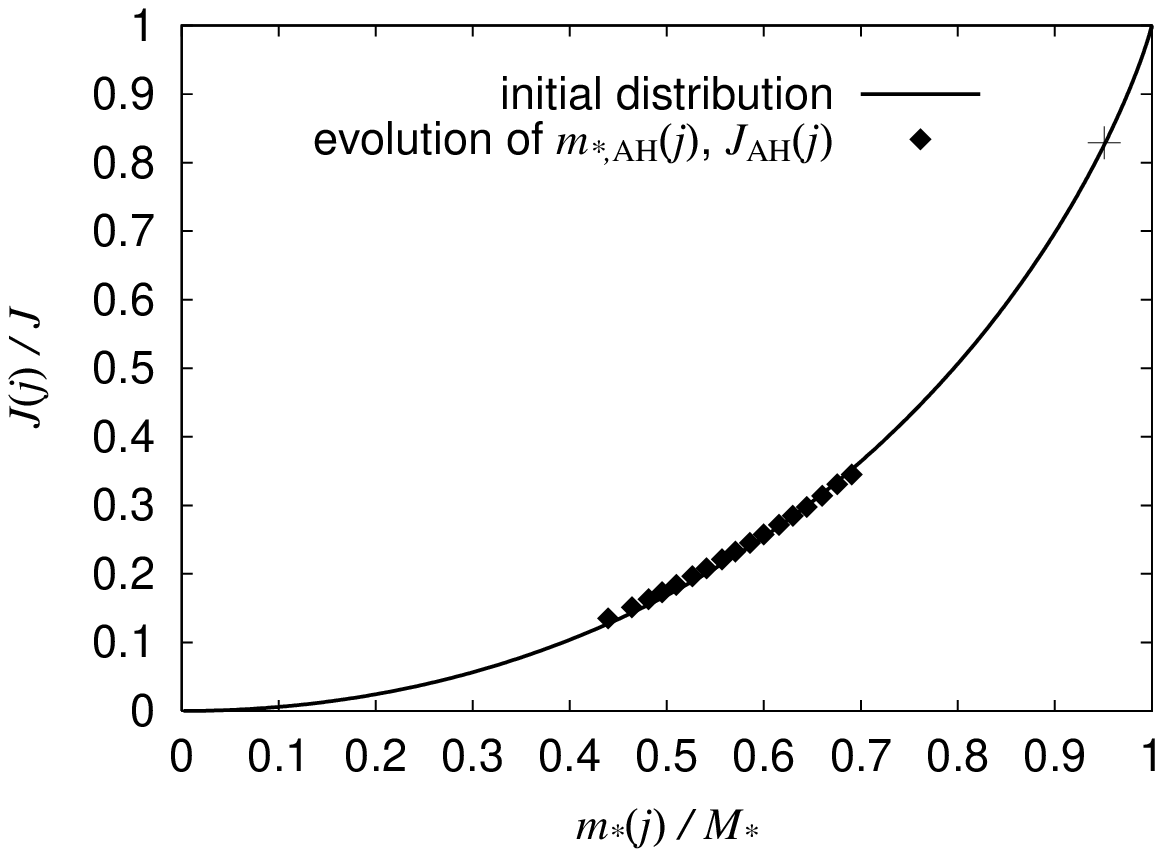}
\end{center}
\caption{Evolution of the baryon rest mass $m_{\ast, \rm AH}(j)$ and the
angular momentum $J_{\rm AH}(j)$ enclosed inside apparent horizons 
(filled diamonds) for the rigid rotation models with 
(a) $\Omega_{0} = 0.065$ and 
(b) $\Omega_{0} = 0.050$. 
The solid curve denotes the initial distribution relation of 
$m_{\ast}(j)$ and $J(j)$. The cross denote the location
of $m_*(\jisom)$ and $J(\jisom)$, which 
are the predicted final values
of the mass and the angular momentum of a black hole. 
}\label{figure12} 
\end{figure}

We perform simulations for various initial models with 
$\Gamma = 1.5$ listed in Table \ref{Table1},
varying grid spacing as $\Delta x = 0.018$,
$0.015$, $0.01125$, $0.009$, and $0.006$ (in units of $G=c=K=1$).
In the simulations, 
the uniform grid is adopted and the outer boundaries along 
the $x$ and $z$-axes are located at $L \sim 10$. 
Numerical simulation is performed on 
FACOM VPP5000 in the data processing center of the National Astronomical
Observatory of Japan, and personal computers with Pentium 4
processors, each of which has 2 Gbytes memory and a 3.0 GHz clock. 

In. Fig. \ref{figure6}(a) and \ref{figure6}(b),
evolution of the central value of $\rho_*$ ($\rho_{*,c}$) and the
central value of the lapse function $\alpha_{c}$ for $\Gamma=1.5$ and 
for the differentially rotation cases with $R_S=R_0/1.5$ and 
$\Omega_{0} = 0.090$, $0.100$, and $0.115$. 
We find that for models with $q_{c} < 1$ ($\Omega_{0} \leq 0.100$),
the collapse proceeds almost monotonically to
form an apparent horizon irrespective
of values of $q$ and the differential rotation parameter $R_{0}$. 
On the other hand, 
for all the models with $\Omega_{0} = 0.115$ for which 
$q_{c} > 1$, we find that a black hole is not formed and
that the stars experience bounces and oscillation. 
This results from the fact that the centrifugal force 
near the axis is too strong to form a seed black hole. 

In Figs. \ref{figure7} and \ref{figure8}, we plot 
snapshots of the density contour curves of $\rho_*$ and
of the velocity field of $(v^x, v^z)$ at selected time slices
for the rigid rotation model with $\Omega_0=0.065$ and
for the differential rotation model with $\Omega=0.115$ and
$R_0=R_S/1.5$.
As we predicted in the last section, the collapse first proceeds
along the rotational axis ($z$-axis) to be a disk-like
structure. Then, the disk collapses to the center to form a
black hole for $q_c < 1$.
Irrespective of the models with $q_c < 1$, the collapse and
the black hole formation proceed in essentially the same manner.
For $q_c > 1$ [see Fig. \ref{figure8}], the centrifugal
force is strong enough to prevent a black hole formation. 
In this case, an oscillating disk is the outcome after the collapse.
Such disk may be unstable against a nonaxisymmetric instability. 

Even if the value of $q$ is much larger than unity, a black hole is formed
for $q_c <1$, as predicted in Sec. III. 
To confirm this conclusion strictly, we performed 
convergence tests varying the grid resolution for a wide range.
In Figs. \ref{figure9}(a) and \ref{figure9}(b), 
we show $H_{\rm error}$ and the
central density $\rho_{*,c}$ with different grid spacing
($\Delta x = 0.009$, 0.015, and 0.018) for 
$\Omega_{0} = 0.065$ and $R_{0} = 1.5R_{S}$. 
These figures illustrate that second-order convergence is achieved.

To accurately determine the threshold values of $q$ or $q_c$, 
it is also crucial that the rest-mass distribution 
$m_{\ast}(j)$ as a function of specific 
angular momentum is conserved accurately at least until
the first formation of an apparent horizon. 
In Fig. \ref{figure10}(a) and \ref{figure10}(b), we 
compare the rest-mass distribution at $t=0$ and at the first
formation of apparent horizon for the rigid rotation case with 
$\Omega_{0} = 0.065$ and $0.090$. 
These figures demonstrate that the rest-mass distribution as
a function of $j$ is conserved well, implying that a spurious
numerical transfer of the angular momentum is negligible. 

In Fig. \ref{figure11}(a) and \ref{figure11}(b),
we show time evolution of the apparent horizon
mass for rigid rotation models with $\Omega_{0} = 0.065$ and $0.090$.
These figures indicate that the process of the black hole formation can be
divided into two phases. One is a phase in which a seed black hole is
formed at the central region. The other is a phase in
which the seed black hole grows as the ambient fluids fall into it. 
We define the mass of the seed apparent horizon $M_{\rm AH, seed}$ 
from the location of the break of the curves for $M_{\rm AH}(t)$. 
{}From this identification, it is found that $M_{\rm AH, seed}$ is much 
smaller than the total mass of the system $M$ and $M_{\rm BH}$
defined in Eq. (\ref{approx-MBH}). 
As a reasonable result, it is also found that 
$M_{\rm AH, seed}$ is smaller for the larger value of $\Omega_{0}$ and
$R_{0}$ (i.e., for the more rapidly and rigidly rotating cases).  

For a Kerr black hole of mass $M_{\rm BH}$ and a spin 
parameter $q_{\rm BH}$, the irreducible mass of the event
horizon $M_{\rm irr}$ is defined as \cite{Singu1} 
\beq \label{def-Mirr}
\left( M_{\rm irr}[M_{\rm BH}, \, q_{\rm BH}] \right)^{2} \equiv  
\frac{1}{2} M_{\rm BH}^{2} \left[
1 + \sqrt{1 - q_{\rm BH}^{2}}
\right] .
\eeq
If the final state of a black hole can be approximated as a Kerr black hole 
even though it is surrounded by an appreciable disk, $M_{\rm AH}$ should
asymptotically approach $M_{\rm irr}[M_{\rm BH}, \, q_{\rm BH}]$. 
Assuming that $M_{\rm BH}$ and $q_{\rm BH}$ may be evaluated from the
approximate relations (\ref{approx-MBH}) and (\ref{approx-JBH}),
$M_{\rm irr}/M$ is $\approx 0.69$ and 0.34 
for the rigidly rotating initial data with 
$\Omega_{0} = 0.065$ and $0.090$, respectively 
(see dotted lines in Fig. \ref{figure11}). We note that 
for the differentially rotating cases, the value is slightly larger, but
it is not significantly different for the same value of $\Omega_0$. 
Thus, $M_{\rm AH}$ should approach these values.
Actually, the value appears to approach them. 
Unfortunately, the computations crash in $\sim 20 M$ after the formation of
apparent horizons, due to the grid stretching around the black hole
horizons, and thus, we cannot confirm this prediction strictly.
To carry out a simulation beyond this time, the so-called 
black hole excision techniques \cite{excision} are necessary. 

It should be noticed that $M_{\rm AH, seed}$ is always smaller than
$M_{\rm irr}$ by a factor of $\sim 2$. This implies that 
the black hole significantly grows during the accretion
phase. For the realistic progenitor of a black hole formation
that have a centrally-condensed density profile, 
the accretion phase is likely to be much longer than 
the collapse timescale (defined as $\rho_{\rm c,0}^{-1/2}$
where $\rho_{\rm c,0}$ is the central density at $t=0$). 
At the accretion phase,
the system is composed of a black
hole and a surrounding massive disk accreting onto the central black hole.
Therefore, 
this accretion phase might be associated with the gamma ray bursts
\cite{GRB}.

For all the models with $\Omega_{0} = 0.100$ in which 
$q_{c} \approx 0.91$--0.96, we find that 
a black hole is formed. However, it is difficult 
to obtain a convergent result with the present computational setting
because of the restriction of the grid resolution. First, 
recall that the grid spacing required to follow a black hole 
formation is $\Delta x \alt 0.1M_{\rm seed, AH}$ 
(not $\Delta x \alt 0.1M$), with which 
a black hole horizon is covered by $\sim 10$ grid points. 
For the model with $\Omega_{0} = 0.100$, we found that 
mass of the seed black hole is $M_{\rm seed, AH} \alt 0.1 M$. 
This implies that for this model, the required grid spacing is
$\Delta x \alt 0.1 M_{\rm seed, AH} \alt 0.01 M \approx 0.005$. 
This value is almost the same as the finest grid spacing we adopt,
indicating it difficult to obtain convergent results.

{}From the above results, we have confirmed 
the conjectures (I), (II), and (IV) 
suggested in Sec. \ref{Prediction}: 
We have found that the quasi-local value $q_{c} \approx 1$
is an approximate threshold for a seed black hole formation and
that after its formation, the black hole grows due to accretion. 

To confirm the conjecture (III), 
we compute the evolution of the
total baryon rest mass and the total angular momentum enclosed
inside an apparent horizon (denoted as $m_{\ast, {\rm AH}}(j)$
and $J_{{\rm AH}}(j)$ respectively) at each time step and compare
their relation with the initial one between $m_{\ast}(j)$ and $J(j)$. 
In Fig. \ref{figure12}(a), we display evolution of 
$m_{\ast, \rm AH}(j)$ and $J_{\rm AH}(j)$ (the diamonds) 
together with the initial relation (the solid curve) for the
rigid rotation model with $\Omega_{0} = 0.065$. 
It is found that an evolutionary track of the black hole 
is approximately determined by the initial distribution
of the mass and the angular momentum, until
a numerical error is accumulated significantly. 
To be more convinced of this result, we also perform an
additional simulation for a moderately rotating star (rigid rotation with 
$\Omega_{0} = 0.050$). 
In Fig. \ref{figure12}(b), we show the evolution track of $m_{\ast}(j)$ and
$J(j)$ together with the initial distribution for this case. This 
figure shows that the initial distribution certainly determines
an approximate evolutionary track of the rest mass and the angular momentum 
enclosed inside the apparent horizon.
These results confirm the conjecture (III)
of the previous section.
This conclusion is quite natural in particular in the present case
because we initially deplete the pressure by a significant factor 
to quickly form a disk-like structure, for which
fluid elements of the same value of a cylindrical radius have
almost the same value of $j$. 

To this time, we have confirmed the conjectures (I) -- (IV). Now, 
let us consider the conjecture (V). To confirm it,
it is necessary to continue simulations for a long time
after formation of apparent horizons. Unfortunately, 
computations crash in $\sim 20 M$ after the formation of
apparent horizons, due to the grid stretching around the black hole
horizons. To carry out a simulation beyond this time
for confirmation of the conjecture (III), the so-called 
black hole excision techniques \cite{excision} are absolutely necessary.

\subsection{Results for $\Gamma = 2.0$}

\begin{figure}[tb]
\begin{center}
\epsfxsize=3.3in
\leavevmode
(a)\epsffile{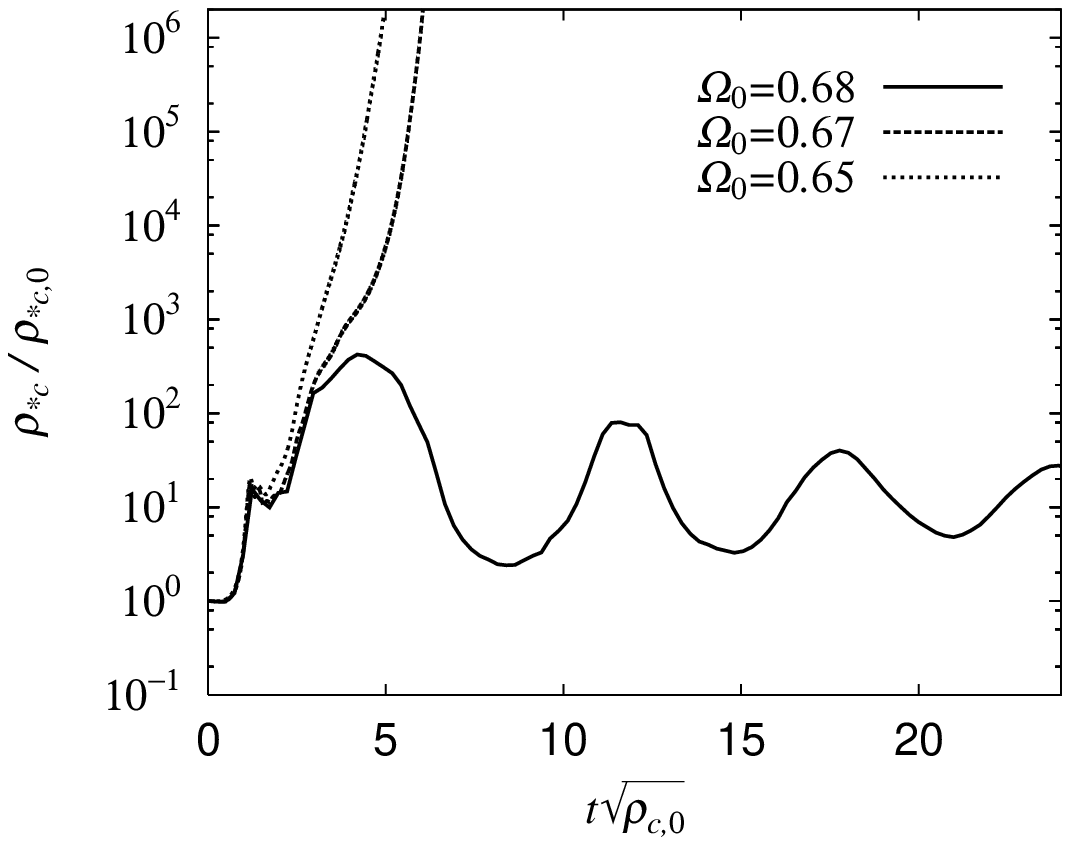}
\epsfxsize=3.3in
\leavevmode
(b)\epsffile{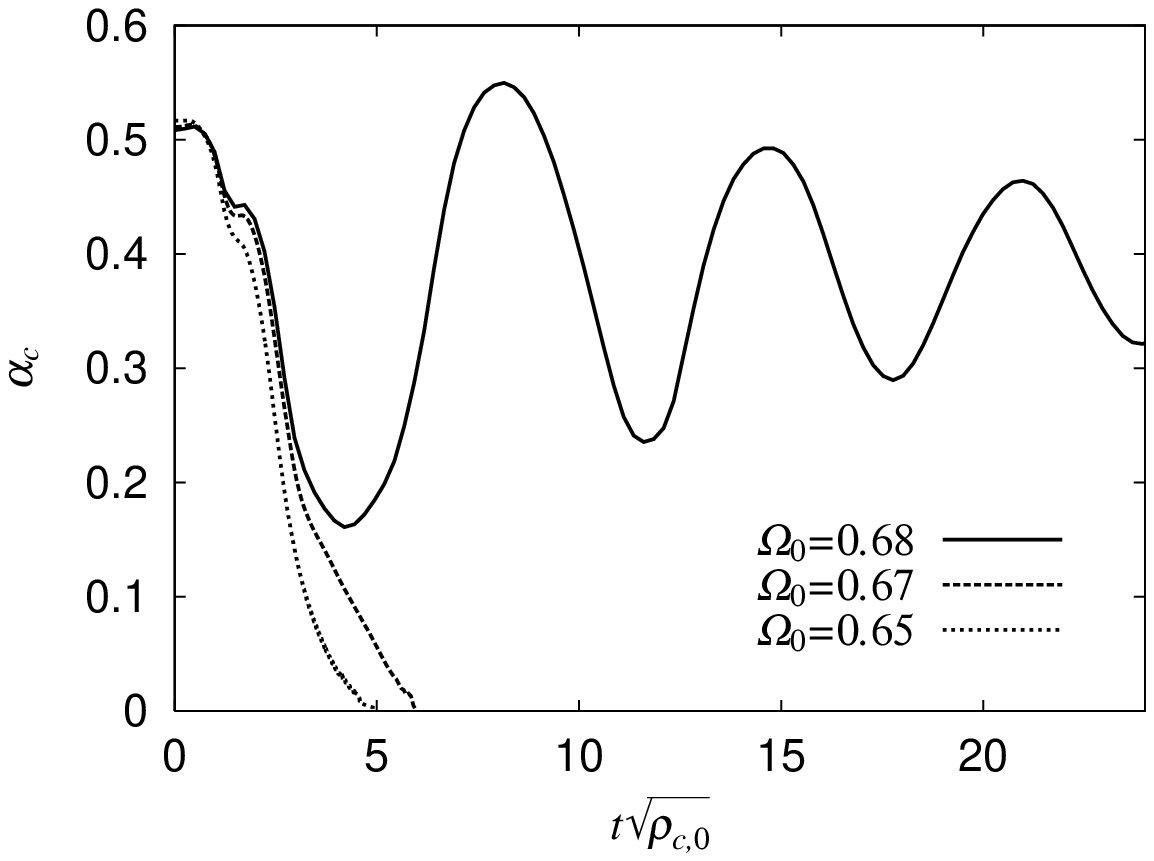}
\end{center}
\caption{Evolution of (a) the central density $\rho_{*c}$ and (b) the
central value of the lapse function $\alpha_{c}$ for $\Gamma=2$. 
In both figures, solid, dashed, and dotted curves denote the models of
$\Omega_{0} = 0.68$, $0.67$, and $0.65$ respectively. 
}\label{figure13}
\end{figure}

For comparison with the $\Gamma=1.5$ case, 
we also study rotating collapse adopting a stiff
equation of state with $\Gamma = 2$. 
The simulations are performed with $N=1000$ and $\Delta x=0.005$
(in units of $G=c=K=1$) for the initial conditions listed in Table II. 

{}From the simulations, we find that a black hole is formed even 
from the initial data sets with $q > 1$. 
In Figs. \ref{figure13}(a) and \ref{figure13}(b), we show time evolution 
of the central density $\rho_{*c}$ and the central value of the lapse
function $\alpha_{c}$ for models listed in Table II. These figures
indicate that the rigidly rotating stars with $\Omega_{0} \alt 0.67$ 
and $q \alt 1.16$ 
collapse to form a black hole. On the other hand, for
rigidly rotating stars with $\Omega_{0} \agt 0.68$ and 
$q \agt 1.17$, a black hole is not formed. Therefore, the threshold value 
of $q$ is between these two values. It should be addressed that the present 
critical value  $q \sim 1.2$ agrees approximately with that found by 
Stark and Piran \cite{SP}. 
For the critical model, the value of $q_{c}$ is $\approx 0.92$, close
to unity. 
Since there is no reason to believe that a particularly large value 
$q \sim 1.2$ should be the
threshold for the black hole formation, we propose the quasi-local value 
$q_{c} \approx 0.92 \sim 1$ as the threshold. 

The plausible reason that the critical value of $q_{c}$ is somewhat
smaller than unity may be as follows. 
First note that since configuration of stars with 
stiffer equations of state is less centrally-condensed and more
uniform, the collapse proceeds in a more coherent 
manner; i.e., the fluid elements collapse to a black hole
rather simultaneously.
Accordingly, whether a black hole is formed or 
not after the collapse is likely to be not completely determined by local 
properties of the stars. Indeed, we find that ratio of the mass of the
seed apparent horizon to the total mass of the system for 
$\Gamma = 2.0$ is much 
larger than that for $\Gamma = 1.5$ case; for $\Gamma = 2.0$, 
$M_{\rm AH, seed}/M \sim 0.35$ at the critical value of $q_{c}$, 
in contrast with that for $\Gamma = 1.5$,
$M_{\rm AH, seed}/M \sim 0.1$. This indicates that for $\Gamma=2$,
not $q_c$ but a value of $q(j)<1$ of a moderately large value of $j$
(denoted as $j_1$) 
may determine a black hole formation. Since $q(j)$ is an increasing
function of $m_{\ast}(j)$, $1 \sim q(j_1) > q_c$. 
Therefore, the threshold value of $q_{c}$ may be
smaller than unity for stiff equations of state. 

Note that for a ``realistic''
progenitor of black hole formation reviewed in Sec. \ref{Candidates},
the associated equations of state are 
soft ($\Gamma \alt 4/3$) ones. Since mass distribution of such stars is 
significantly centrally-condensed, a demarcation between complete
collapse and a bounce will be determined by properties of the central
region. Consequently the critical value of $q_{c}$ may be almost
unity. Furthermore, since the ratio $M_{\ast}/M$ is almost unity for
this case, $q_{\ast, c} \approx 1$ may be also used as the threshold for
black hole formation. 

Before closing this section, let us explain the reason
that $q \approx 1$ has been believed as a threshold for
black hole formation in the previous works. Stark and Piran used  
a stiff ($\Gamma = 2$) equation of state \cite{SP} as we do. 
As described above, such stars have rather uniform density 
distribution and the collapse proceeds in a coherent manner. 
Furthermore, the distribution of $q(j)$ expressed as a function of
$m_{\ast}(j)$ inside stars with stiff equations of state is rather flat. 
Indeed, for $\Gamma = 2.0$ and rigidly rotating case, the ratio 
of $q$ to $q_c$ is $\sim 1.25$ in contrast with the 
$\Gamma = 1.5$ cases for which the ratio is $\sim 2$
(compare Fig. \ref{figure5} and Figs. \ref{figure3}(a) for rigid
rotation models). 
These two facts explain the reason that Stark and Piran concluded that 
$q \approx 1$ is an approximate threshold for black hole formation. 

Nakamura and his collaborators performed simulation for highly
differentially rotating stars \cite{Nak,NOK}. 
The higher the degree of differential rotation becomes, the flatter 
the distribution of $q(j)$ is, as indicated in
Fig. \ref{figure4}. As a result, the difference between $q$ and $q_c$ 
becomes small for their initial data.
This is the reason that the value $q=1$ was regarded
as the demarcation between black hole formation and a bounce. Abrahams
{\it et al}.\cite{ACST} performed simulations for toroidal star 
clusters. Having a toroidal configuration implies
that they are highly differentially rotating. 
Thus, the value of $q_{c}$ is likely as large as $q$, and hence, the
global parameter $q$ may be used as the threshold. 

\section{SUMMARY}

We have reported new results about black hole formation in collapse
of rapidly rotating stars with $q > 1$ and with 
moderately soft ($\Gamma = 1.5$) and stiff ($\Gamma = 2.0$) equations of
state, analyzing initial conditions and performing 
axisymmetric simulations in full general relativity. 

The initial conditions are given, following Stark and Piran \cite{SP}; 
we first give marginally stable spherical 
polytropes with $\Gamma = 1.5$ or $\Gamma = 2.0$ and then 
artificially add an angular momentum. 
Adopting the same method described in \cite{ShibaSA}, we have predicted 
that (I) inner region in which $q_{c} \alt 1$ will collapse first to form
a seed black hole even if the global value of $q$ is much larger than
unity; (II) the formed seed black hole will subsequently grow
as the ambient fluids accrete onto it; 
(III) evolution of the relation between 
the rest mass $m_{\ast}(j)$ and angular momentum $J(j)$ enclosed
inside a growing black hole will agree approximately with
the initial relation between $m_{\ast}(j)$ and $J(j)$; 
(IV) whether black hole is formed or not can be determined
by $q_{c}$ irrespective of $q$; 
(V) the final outcome of dynamical
collapse will be a black hole surrounded by an appreciable disk.  

To confirm these conjectures, we performed fully general relativistic
hydrodynamic simulations using a high-resolution
shock capturing scheme with the $\Gamma$-law equations of state. 
As a result of numerical simulations, we confirmed the conjectures (I),
(II), (III), and (IV).
{}From these results we conclude that the previous criterion for black 
hole formation that no black hole is formed for $q \agt 1$ is no longer
valid in particular for soft equations of state. 
As a new criterion for no black hole formation, we propose the condition 
$q_{c} \agt 1$. 

Associated with conjecture (II), we have found that the process of
black hole formation is divided into two phases:
the first phase is that a seed black hole is formed
and the other is the accretion phase that a part of the ambient fluids 
are swallowed into the formed seed black hole. At the time of
the accretion phase, the system is composed of a black
hole and a surrounding massive disk accreting onto the central black hole. 
In the collapse of massive iron cores, 
such accretion phase may be associated with the gamma ray bursts \cite{GRB}. 

In this paper, we have focused only on the criterion of black hole formation  
analyzing simple toy models. To obtain a scientific result 
that can be compared with observational data such as gravitational waves,
however, a simulation with a realistic initial condition and
a realistic equation of state is necessary. As we indicate in this paper,
a rapidly rotating black hole will be formed after rapidly
rotating massive stellar core collapse and pair-unstable collapse, 
if the progenitors are not extremely rapidly rotating
at the onset of collapse as $q_c < 1$. 
Formation of a rapidly rotating black hole with $q \sim 1$ and
accretion of a large mass onto such black hole of a large value of
$q$ are likely to be strong burst sources of gravitational waves for
laser interferometric detectors. 
Thus, performing realistic numerical simulations of black hole formation 
is an important subject for predicting the gravitational waveforms. 
We plan to attack these computations in a fully general relativistic
manner extending previous works \cite{S-Sekig1,S-Sekig2}. 

For $q_c > 1$, a black hole will not be formed promptly.
In such case, the collapse leads to formation of a self-gravitating
disk or torus. They 
will be subsequently unstable against nonaxisymmetric 
deformation. After the nonaxisymmetric instabilities turn on,
angular momentum will be transported from the inner region to
the outer region, decreasing the value of $q(j)$ around the inner region
below unity. As a result, a seed black hole may be formed. 
To follow these processes, it is necessary to perform a numerical
simulation without assuming the axial symmetry.
During the collapse, the typical length scale may change 
by a factor of $10^4$ from $R / M \sim 10^4$ to 1. 
To follow the collapse by numerical simulation, 
very large computational resources will be necessary and, thus, 
the simulation for this phenomenon will be one of 
the computational challenges in the field of numerical relativity.

\vspace{4mm}

\begin{center}
{\bf Acknowledgments}
\end{center}

Numerical computations were performed 
on the FACOM VPP5000 machine in the data processing center of 
National Astronomical Observatory of Japan.
This work is in part supported by Japanese Monbu-Kagakusho Grants 
(Nos. 14047207, 15037204, and 15740142). 

\end{document}